\documentclass[aps,pra,twocolumn,superscriptaddress,amsmath]{revtex4-2}
\usepackage{hyperref,orcidlink,amsmath,amssymb}
\hypersetup{colorlinks=true, citecolor=blue, urlcolor=blue, linkcolor=blue} 
\newenvironment{proof}[1][Proof]{\noindent\textbf{#1.} }{\ \rule{0.5em}{0.5em}}

\begin{document}
\title{Chiral quantum batteries}
\author{Rong-Fang Liu}
\affiliation{Key Laboratory of Quantum Theory and Applications of Ministry of Education, Lanzhou Center for Theoretical Physics, Gansu Provincial Research Center for Basic Disciplines of Quantum Physics, Lanzhou University, Lanzhou 730000, China}
\affiliation{Key Laboratory of Time Reference and Applications, Chinese Academy of Sciences,
Innovation Academy for Precision Measurement Science and Technology, Chinese Academy of Sciences, Wuhan 430071, China}
\author{Wan-Lu Song\orcidlink{0000-0002-6437-9748}}
\affiliation{Department of Physics, Hubei University, Wuhan 430062, China}
\author{Wan-Li Yang}
\email{ywl@wipm.ac.cn}
\affiliation{Key Laboratory of Time Reference and Applications, Chinese Academy of Sciences, Innovation Academy for Precision Measurement Science and Technology, Chinese Academy of Sciences, Wuhan 430071, China}
\author{Hua Guan}
\email{guanhua@apm.ac.cn}
\affiliation{Key Laboratory of Time Reference and Applications, Chinese Academy of Sciences,
Innovation Academy for Precision Measurement Science and Technology, Chinese Academy of Sciences, Wuhan 430071, China}
\author{Jun-Hong An\orcidlink{0000-0002-3475-0729}}
\email{anjhong@lzu.edu.cn}
\affiliation{Key Laboratory of Quantum Theory and Applications of Ministry of Education, Lanzhou Center for Theoretical Physics, Gansu Provincial Research Center for Basic Disciplines of Quantum Physics, Lanzhou University, Lanzhou 730000, China}

\begin{abstract}
 Quantum batteries (QBs) offer compelling advantages over conventional ones in terms of superior energy density, ultrafast charging, and high conversion efficiency. However, their realization is hampered by decoherence, which causes incomplete charging, rapid self-discharging, and reduced extractable work. Here, we propose a quantum-compatible battery architecture based on a chiral magnonic platform. It comprises two yttrium iron garnet spheres, one serving as the charger and the other as the QB, coupled to a waveguide. The unique chiral coupling between magnons and the guided electromagnetic fields breaks inversion symmetry, inducing both nonreciprocal energy flow and coherent interference between the charger and QB. Their synergy endows our QB with a 34-fold increase in energy capacity and a 55-fold boost in extractable work compared to its achiral counterpart in an experimentally accessible regime. Our scheme harnesses the decoherence induced by the common electromagnetic fields, which enables the robustness and wirelesslike remote charging features of the QB. Without relying on quantum resource and being representable by classical dynamics, our architecture provides a viable route toward the realization of robust and remote-charging QBs.
\end{abstract}
\maketitle

\section{Introduction}
The growing demand for energy resulting from industrial expansion, population growth and the transition to renewable and miniaturized energy systems makes it urgent to transform conventional energy storage technologies. Quantum batteries (QBs) are an innovative technology that could revolutionize existing battery technologies by using quantum mechanical principles, such as quantum superposition, entanglement, and coherence, in the storage and release of energy \cite{PhysRevE.87.042123,RevModPhys.96.031001}. They exhibit remarkable superiority to conventional batteries in terms of increased energy storage capacity \cite{PhysRevResearch.4.043150,PhysRevLett.131.030402,PhysRevA.109.042424}, faster charging rates \cite{PhysRevLett.118.150601,PhysRevLett.120.117702,PhysRevLett.125.236402,PhysRevLett.128.140501}, and higher extractable-work efficiency \cite{PhysRevLett.122.047702,PhysRevLett.122.210601,PhysRevLett.124.130601,PhysRevLett.129.130602,PhysRevLett.131.060402}. The proof-of-principle realizations of QBs have been demonstrated in various platforms, including quantum dots \cite{PhysRevLett.131.260401}, trapped ions \cite{PhysRevLett.135.140403}, linear optics systems \cite{PhysRevLett.131.240401}, and organic microcavities \cite{Quach2022}. 

The realization and application of QBs are hindered by ubiquitous decoherence in the microscopic world. Decoherence degrades the charging efficiency, induces spontaneous energy loss called self-discharging, and reduces the extractable work of QBs. Many strategies have been proposed to overcome these unwanted effects caused by decoherence. It has been found that the ideal efficiency of QBs is recovered using remote charging mediated by a waveguide \cite{PhysRevLett.132.090401,fn1b-2m9g,c4hd-8vfj,fwdy-lccn}, topological phases of the reservoir \cite{PhysRevLett.134.180401,Cavazzoni2026}, nonreciprocal charging \cite{PhysRevLett.132.210402,p93y-jflt,67wh-1fxv,PhysRevA.112.022214}, non-Hermitian exceptional points \cite{PhysRevA.109.042207,PhysRevA.112.042205}, non-Markovian effects \cite{Kamin_2020}, feedback control \cite{Mitchison2021chargingquantum}, and noise-assisted charging \cite{PhysRevLett.122.210601,PhysRevA.104.032207,PhysRevE.104.044116,PhysRevLett.134.130401,vqnk-kzqg}. Further studies reveal that the self-discharging of QBs is suppressed by Floquet engineering \cite{PhysRevA.102.060201} and the enhancement of quantum coherence via hyperfine interactions in nitrogen-vacancy centers \cite{PhysRevLett.135.020405}. Therefore, in the current noisy intermediate-scale quantum era, developing more stable QB architectures with intrinsic robustness to decoherence and uncovering more useful quantum resources to enhance the QB performance in the decoherence situation are highly desirable \cite{Ferraro2026}. On the other hand, there is an increasing interest in studying the chiral degrees of freedom in different light-matter interfaces \cite{doi:10.1126/science.1257671,PhysRevLett.113.237203,sollner2015deterministic,lodahl2017chiral,owens2022chiral,science.abq7870,Natphoton2025,PhysRevLett.127.233601,PhysRevLett.134.173601,doi:10.1126/science.aan8010,PhysRevResearch.2.023003,PhysRevLett.132.223201,PhysRevLett.134.013602}. It has become a powerful resource for engineering quantized light-matter interactions, enabling the development of chiral quantum control in giant atom systems \cite{PhysRevLett.126.043602,PhysRevLett.133.063603,Chang2025,Chen2022} and quantum interconnects in chiral waveguide- or cavity-QED systems \cite{Almanakly2025,PhysRevA.108.013507,PhysRevA.94.012302,PRXQuantum.3.020363}.
A natural question is whether the chirality of quantum systems can be used to enhance the QB performance in the presence of decoherence. 
\begin{figure}
\includegraphics[width=0.9\columnwidth]{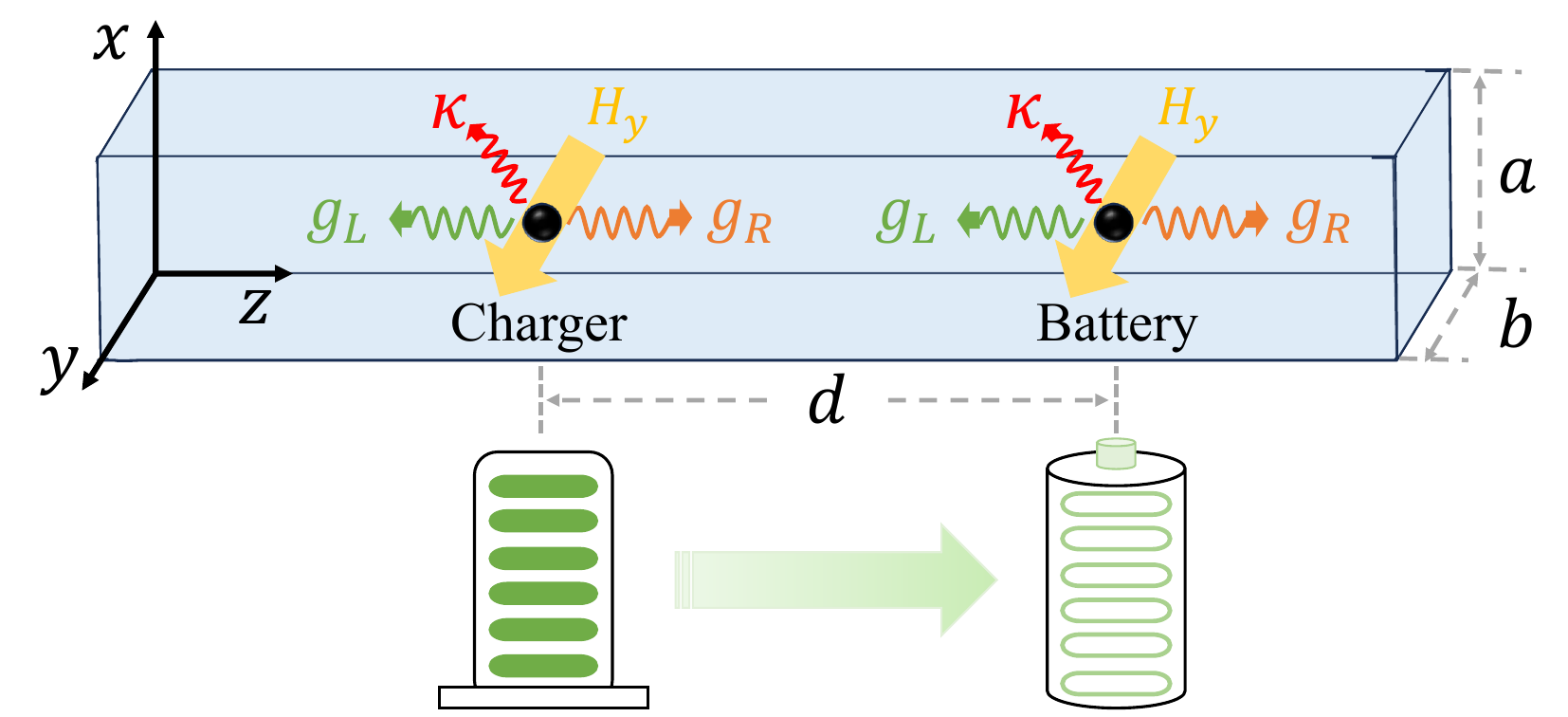}
\caption{Schematic of a chiral QB. Two YIG spheres separated by a distance $d$ are coupled to a waveguide via the chiral coupling rates $g_{L/R}$ along the $z$-axis. A static $y$-axis magnetic field $\mathcal{H}_{y}$ is applied to produce an eigenfrequency of the magnons in the YIG spheres. $\kappa$ is the damping rate of the magnons. The left YIG sphere acts as the charger and the right one acts as the QB.}\label{moxingtu}
\end{figure}

Based on the fact that the eigenenergies of continuous-variable systems are unbounded from above, it is expected that such systems have a greater energy capacity than the discrete-variable systems. Inspired by the recent realization of chiral magnon-photon coupling in a spin-momentum locked waveguide \cite{Qian2025}, we propose a continuous-variable QB scheme consisting of a pair of distant magnons that are chirally coupled to the electromagnetic fields (EMFs) in a waveguide. The chiral interactions result in the breaking of the inversion symmetry, leading to nonreciprocal energy flow and unbalanced coherent interference between the charger and QB. Their interplay enables our QB to exhibit an enhancement of 2 orders of magnitude in both the energy capacity and the extractable work compared to its achiral counterpart, even in the presence of local magnon dissipation. In particular, our QB works flawlessly in a wirelesslike remote-charging manner. Our results indicate that chiral magnon-waveguide coupling provides a viable mechanism for improving the charging performance of QBs in practical situations.

The remainder of this paper is organized as follows. In Sec. \ref{sec:model}, we give our chiral QB scheme. In Sec. \ref{nmrcer}, we present numerical results on the chirality enhanced performance of our QB scheme. In Sec. \ref{scalbst}, we generalize our scheme to $N$-cell case to demonstrate its scalability. Finally, we summarize our main findings and conclusions in Sec. \ref{sec:conclusion}.

\section{Chiral QB scheme}\label{sec:model}
We propose a chiral QB scheme. It consists of two YIG spheres embedded in a rectangular hollow metallic waveguide (see Fig. \ref{moxingtu}). Having a common frequency $\omega_0 = \gamma \mathcal{H}_{y}$, with $\gamma /2\pi= 28$ GHz/T being the gyromagnetic ratio and $\mathcal{H}_{y}$ being the static $y$-axis field, the magnons generated in the spheres form a continuous-variable bimodal system. The magnon of the left YIG sphere serves as the charger and the right one serves as the QB, both of which are initially in the vacuum state. The energy is supplied to the charger by a driving field. The QB and the charger are far separated such that the energy cannot be coherently transferred to the QB. This design can prevent the destruction of the local decoherence of the charger and QB on the charging. The Hamiltonian is $\hat{H}(t)=\hat{H}_{\text{m}}(t)+\hat{H}_{\text{w}}+\hat{H}_{\text{i}}$, with ($\hbar=1$):
\begin{eqnarray}\label{Eq1}
\hat{H}_{\text{m}}(t)&=&\sum_{j=1,2}\omega_0(\hat{m}_{j}^{\dagger}\hat{m}_{j}+\tfrac{1}{2})
+\Omega(\hat{m}_{1}^{\dagger}e^{-i\omega_d t}+\text{H.c.}),\\
\hat{H}_{\text{w}}&=&\sum_{\lambda=L,R}\int\omega\hat{b}_{\lambda}^{\dagger}(\omega)\hat{b}_{\lambda}(\omega)d\omega,\\
\hat{H}_{\text{i}}&=&\sum_{\substack{\lambda=L,R\\j=1,2}}\int d\omega
[{ig_{\lambda j}(\omega)\over\sqrt{2\pi}}\hat{b}_{\lambda}^{\dagger}(\omega)\hat{m}_{j}e^{-i{\omega \over v_{\lambda}}z_{j}}+\text{H.c.}],~~~~
\end{eqnarray}
where $\Omega$ and $\omega_d$ are the amplitude and frequency of the driving field, $\hat{m}_{j}$ and ${\bf r}_{j}=(x_{j},z_{j})$ are the annihilation operators and the positions of the charger and QB, and $\hat{b}_{L/R}(\omega)$ are the annihilation operators of the left- and right-propagating EMFs with frequency $\omega$ and group velocities $v_{L/R}$ in the waveguide. The coupling strength between the $j$th magnon and the EMFs is $g_{\lambda j}(\omega)=\mu_{0}\sqrt{\frac{\gamma MV_{j}}{2}}(iB_{x_j}^{\lambda}-B_{z_j}^{\lambda})$, where $B^{\lambda}_{x_j/z_j}$ is the magnetic-field strength felt by the $j$th YIG sphere, $\mu_{0}$ is the vacuum permeability, $M$ and $V_{j}$ are the magnetization strength and the volume of the $j$th YIG sphere. For the $\text{TE}_{10}$ mode, we have $g_{\lambda j}(\omega)=\sqrt{\frac{\gamma MV_{j}}{2\epsilon_{0}\omega ab}}[\frac{\pi}{a}\cos(\frac{\pi x_{j}}{a})-k_{\lambda}\sin(\frac{\pi x_{j}}{a})]$ \cite{PhysRevLett.124.107202,PhysRevB.101.094414}, where $a$ and $b$ are the lengths of the cross section of the waveguide, $\epsilon_{0}$ is the vacuum permittivity, and $k_\lambda=\omega/v_\lambda$ is the wave vector. We consider that the two YIG spheres have $x_1=x_2\equiv x_0$ and $V_1=V_2\equiv V$, which reduce $g_{\lambda1}(\omega)=g_{\lambda2}(\omega)\equiv g_{\lambda}(\omega)$. The relation $k_{R}=-k_{L}\equiv k$ leads to $g_{L}(\omega)\ne g_{R}(\omega)$ and thus the chiral feature of the magnon-EMF coupling. The chirality arises from the elliptically polarized magnetic components, giving rise to the so-called spin-momentum locking \cite{lodahl2017chiral}. We can tune $g_{\lambda j}$ and chirality by adjusting the position $x_{0}$.

Tracing out the EMFs and the local thermal reservoirs from the unitary dynamics of the total system, we derive a Born-Markov master equation satisfied by the density matrix $\rho(t)$ of the charger and the QB as Appendix \ref{Appendix:Master equation}
\begin{eqnarray}\label{Eq2}
\dot{\rho}(t)&=&-i[\hat{H}_\text{m}(t)+\sum_{\lambda=L,R}\hat{H}_{\lambda},\rho(t)]+\big\{\sum_{\lambda=L,R}\Gamma_{\lambda}\check{\mathcal{L}}_{\hat{M}_{\lambda}}\nonumber\\
&&+\sum_{j=1,2}\kappa[(\bar{n}+1)\check{\mathcal{L}}_{\hat{m}_{j}}+\bar{n}\check{\mathcal{L}}_{\hat{m}_{j}^{\dagger}}]\big\}\rho(t),
\end{eqnarray}
where $\Gamma_{\lambda}=g_{\lambda}(\omega_0)^{2}$, $\kappa$, and $\bar{n}=(e^{\omega_{0}/(k_{B}T)}-1)^{-1}$ are the decay rate and thermal excitation number of the two magnons at temperature $T$, $\check{\mathcal{L}}_{\hat{o}}\cdot = \hat{o}\cdot\hat{o}^\dagger - \frac{1}{2}\{\hat{o}^\dagger\hat{o}, \cdot\}$ is the Lindblad superoperator, and $k_{B}$ is the Boltzmann constant. The Hamiltonians $\hat{H}_{L}=-\frac{i\Gamma_{L}}{2}(\hat{m}_{1}^{\dagger}\hat{m}_{2}e^{ik_0d}-\text{H.c.})$ and $\hat{H}_{R}=-\frac{i\Gamma_{R}}{2}(\hat{m}_{2}^{\dagger}\hat{m}_{1}e^{ik_0d}-\text{H.c.})$ describe the coherent charger-QB couplings mediated by the left- and right-propagating EMFs, with $k_0=\omega_0/v_R$ and $d \equiv z_2 - z_1$ \cite{PhysRevB.106.104432}. The operators $\hat{M}_L=\hat{m}_{1}+\hat{m}_{2}e^{ik_{0}d}$ and $\hat{M}_R=\hat{m}_{1}+\hat{m}_{2}e^{-ik_{0}d}$ describe the collective dissipative operators induced by the EMFs. The chirality due to $g_L(\omega)\neq g_R(\omega)$ breaks the inversion symmetry in Eq. \eqref{Eq2}, which is expected to trigger a unidirectional energy flow from the charger to the QB. This nonreciprocity is quantified by \cite{Qian2025}
\begin{equation}
   D = (\Gamma_R - \Gamma_L)/(\Gamma_R + \Gamma_L),
\end{equation} 
where $D = 1$ indicates the fully nonreciprocal case and $D=0$ denotes the reciprocal case. For the parametrization used below, $\Gamma_R$ is fixed as a reference coupling, while $\Gamma_L=\Gamma_R(1-D)/(1+D)$ is varied such that $D$ ranges from $0$ to $1$. Our nonreciprocity stems from the chiral magnon-EMF interaction. It essentially differs from the one in Ref. \cite{PhysRevLett.132.210402}, which comes from the balance between the QB-charger coherent couplings and their local dissipation. 

Our QB is characterized by the following quantities. The first is the stored energy $\mathcal{E}_\text{B}(t)=\text{Tr}[\hat{H}_\text{B}\rho_2(t)]$ with $\hat{H}_\text{B}=\omega_0(\hat{m}_2^\dag\hat{m}_2+\tfrac{1}{2})$ and $\rho_2(t)=\text{Tr}_1[\rho(t)]$ \cite{Allahverdyan2004,PhysRevB.99.035421,PhysRevLett.130.210401}. It quantifies the amount of energy stored in the QB. According to the second law of thermodynamics, not all of $\mathcal{E}_\text{B}(t)$ is convertible into work by unitary transformation \cite{PhysRevLett.124.130601}. Thus, we need a quantity to measure the maximum work that can be extracted from the QB by unitary transformation, i.e., the ergotropy $\mathcal{W}(t)= \mathcal{E}_\text{B}(t)-\text{Tr}[\hat{H}_\text{B}{\tilde\rho}_2(t)]$ \cite{PhysRevLett.125.180603,PhysRevE.102.042111}. Here ${\tilde\rho}_2(t)$ is the so-called passive state whose energy is not unitarily extractable \cite{PhysRevB.99.035421,PhysRevE.87.042123}. For the continuous-variable system, the Gibbs state has been proven to be the passive state. The work-extraction efficiency reads $R(t) \equiv \mathcal{W}(t) / \mathcal{E}_\text{B}(t)$, which describes the fraction of energy that is convertible to useful work. The third quantity is quantum coherence $C(t)=\text{Tr}(\rho_2\log_{2}\rho_2)-\text{Tr}(\rho_2\log_{2}\rho_\text{d})$, where $\rho_{\text{d}}$ is the state obtained from $\rho_2$ by deleting all off-diagonal elements \cite{PhysRevLett.125.180603,PhysRevLett.113.140401}. Quantum coherence plays a constructive role in extracting work \cite{PhysRevLett.125.180603,PhysRevLett.129.130602} and mitigating self-discharging \cite{PhysRevLett.135.020405} of discrete-variable QBs. When $\rho_2$ is Gaussian, we have \cite{PhysRevLett.127.210601,PhysRevA.93.032111}
\begin{eqnarray}
\mathcal{E}_\text{B}(t)&=&\omega_0\{{\text{Tr}[{\pmb\sigma}^{(2)}(t)]/ 4}+{|{\bf d}(t)|^2/ 2}\},\label{EqEB}\\ 
\mathcal{W}(t)&=&\mathcal{E}_\text{B}-\omega_0{\sqrt{\text{det}[{\pmb\sigma}^{(2)}(t)]}}/{2},\label{EqW}\\
C(t)&=&\mathcal{S}(\bar{N})-\mathcal{S}((\sqrt{\text{det}[{\pmb\sigma}^{(2)}(t)]}-1)/2).\label{EqC}
\end{eqnarray}
Here ${\pmb\sigma}^{(2)}(t)$ and ${\bf d}(t)$ are the covariance matrix and displacement vector of $\rho_2(t)$ whose elements are $\sigma_{lm}^{(2)}=\text{Tr}[\{\Delta \hat{r}_{l},\Delta \hat{r}_{m}\}\rho_2(t)]$ and $d_{l}=\text{Tr}[\hat{r}_{l}\rho_2(t)]$, with $\Delta\hat{r}_{l}=\hat{r}_{l}-d_{l}$ and $\hat{r}=({\hat{m}_{2}+\hat{m}_{2}^{\dagger}\over\sqrt{2}},{\hat{m}_{2}-\hat{m}_{2}^{\dagger}\over i\sqrt{2}})^{\text{T}}$, $\mathcal{S}(x)=(x+1)\log_{2}(x+1)-x\log_{2}x$, and $\bar{N}=(\text{Tr}[{\pmb\sigma}^{(2)}(t)]+2|{\bf d}(t)|^{2}-2)/4$. The last quantity is the remote-charging efficiency $\eta(t)=\mathcal{E}_\text{B}(t)/\mathcal{E}_\text{C}(t)$, where $\mathcal{E}_\text{C}(t)=\text{Tr}[\hat{H}_\text{C}\rho_1(t)]$ is the charger energy with $\hat{H}_\text{C}=\omega_0(\hat{m}_1^\dag\hat{m}_1+\frac{1}{2})$ and $\rho_1(t)=\text{Tr}_2[\rho(t)]$. It measures the fraction of the energy that is remotely transferred from the charger to the QB. Solving Eq. \eqref{Eq2} under the resonant condition $\omega_{d}=\omega_{0}$, we derive the steady-state energies and ergotropy, see Appendixes \ref{Dynamical solution} and \ref{Appendix:Energies and ergotropy}, as
\begin{align}
\mathcal{E}_\text{C,ss}&= \omega_{0}\Big[{4\Omega^{2}\alpha^{2}}{|\zeta|^{-2}}+\frac{\kappa\bar{n}\alpha(\xi+2\Gamma_{L}^{2})}{\xi^{2}-4\Gamma_{L}^{2}\Gamma_{R}^{2}}+\frac{1}{2}\Big], \label{Ecinfty}\\ 
\mathcal{E}_\text{B,ss} &=\omega_{0}\Big[{16\Omega^{2}\Gamma_{R}^{2}}{|\zeta|^{-2}}+\frac{\kappa\bar{n}\alpha(\xi+2\Gamma_{R}^{2})}{\xi^{2}-4\Gamma_{L}^{2}\Gamma_{R}^{2}}+\frac{1}{2}\Big],\label{Ebinfty}\\ 
\mathcal{W}_\text{ss}&={16\omega_{0}\Omega^{2}\Gamma_{R}^{2}}{|\zeta|^{-2}},\label{winfty}
\end{align}
where $\alpha=\Gamma_{L}+\Gamma_{R}+\kappa$, $\zeta=\alpha^{2}-4\Gamma_{L}\Gamma_{R}e^{2ik_{0}d}$, and $\xi=\alpha^{2}-2\Gamma_{L}\Gamma_{R}\cos(2k_{0}d)$. The results reveal that, although there is no direct charger-QB interaction, a wirelesslike remote charging from the charger to the QB is realized by the $L/R$-propagating EMFs in the waveguide via their chiral couplings. It is remarkable to find that $\mathcal{E}_\text{B,ss}$ is equal to $\mathcal{W}_\text{ss}$ except for the vacuum zero-point energy when $\bar{n}=0$. Thus, all the energy remotely charged in our QB is extractable to useful work at zero temperature. This is understandable given that the passive state is just the vacuum state for $T=0$. For the parametrization specified above, where $\Gamma_R$ is fixed and $\Gamma_L=\Gamma_R(1-D)/(1+D)$, Eq. \eqref{winfty} gives $\max_{d}{\frac{\mathcal{W}_{\text{ss}|D=1}}{\mathcal{W}_{\text{ss}|D=0}}}\approx 64 $ when $\kappa/\Gamma_{R}\ll 1$. Thus, the extractable work can be enhanced by up to approximately $64$ times relative to the achiral case. For comparison, Ref.~\cite{PhysRevLett.132.210402} reported a fourfold enhancement in the QB energy when comparing the nonreciprocal and reciprocal cases, whereas our scheme yields an approximately $64$-fold enhancement in the extractable work between the chiral and achiral cases.

\begin{figure}[tbp]
\includegraphics[width=0.9\columnwidth]{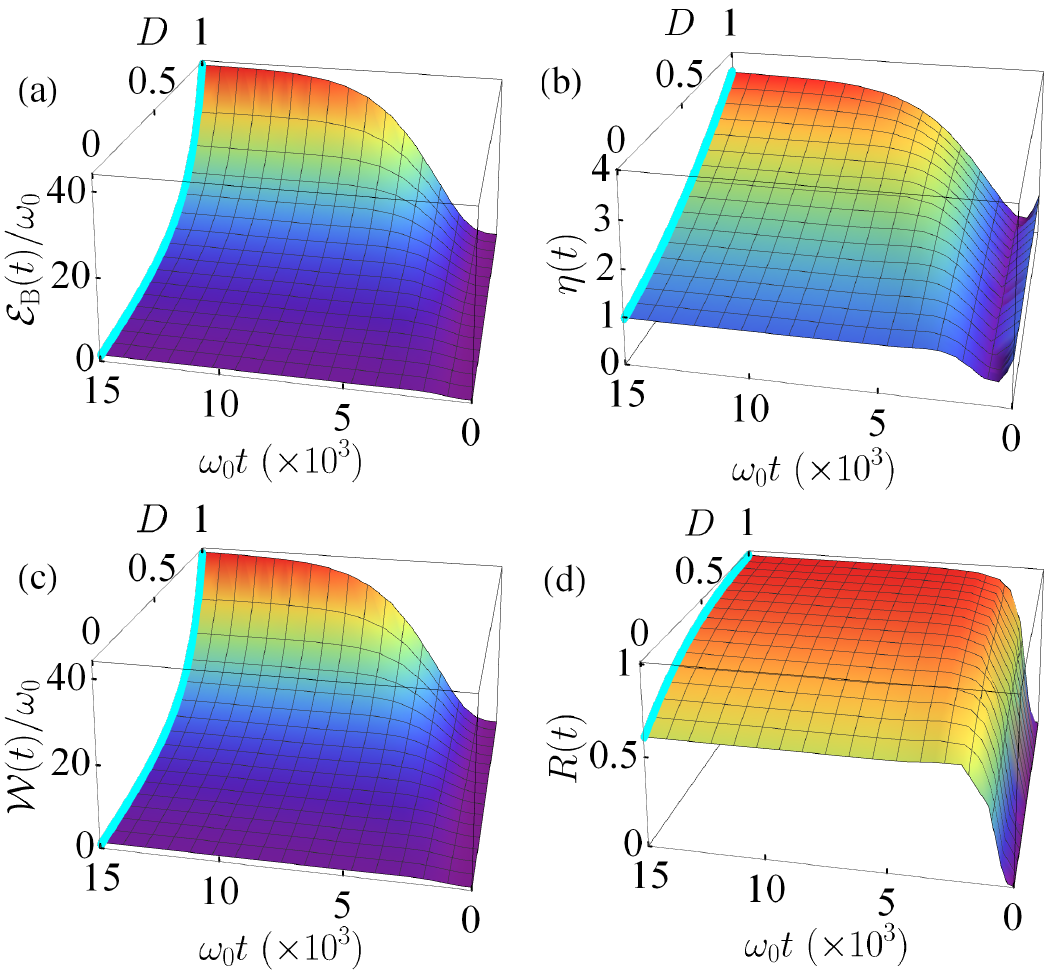}
\caption{Evolution of (a) the stored energy $\mathcal{E}_\text{B}(t)$, (b) the charging efficiency $\eta(t)$, (c) the ergotropy, and (d) the work-extraction efficiency $R(t)$ as a function of the nonreciprocity parameter $D$ when $\Gamma_L$ varies from zero to $\Gamma_R$. The cyan solid line is the steady-state analytical solution. We use $k_0d=\pi/2$, $\omega_d=\omega_0$, $\Omega\approx0.0022\omega_0$, $\Gamma_R\approx0.001\omega_0$, $\bar{n}=0$, and $\kappa=\frac{\Gamma_R}{20}$.}\label{energy}
\end{figure}
\section{Numerical results}\label{nmrcer} Based on experimental conditions of the magnon-waveguide setup \cite{WOS000503012600001,PhysRevB.101.094414,PhysRevLett.124.107202,kang2025magnonicentanglementchiralcavitymagnon}, we choose $\omega_{0}/2\pi=16.2$ GHz, $\Gamma_{R}/2\pi=20$ MHz, $\kappa/2\pi=1$ MHz, $\Omega/2\pi=36$ MHz, and $T = 30$ mK at which $\bar{n}\approx0$. We first focus on how the nonreciprocity governs the charging dynamics by investigating the QB energy $\mathcal{E}_\text{B}(t)$ and the remote-charging efficiency $\eta(t)$. Figure~\ref{energy}(a) shows that, although the QB and charger experience local decoherence at a rate $\kappa$, stable QB energy is still achieved. It indicates that the EMFs as a common environment succeed in mediating a stable energy transfer from the charger to the QB, with the destructive impact of their local decoherence sufficiently suppressed. In contrast to the previous non-Markovian QB scheme \cite{PhysRevLett.132.090401}, our wirelesslike remote-charging QB works under the Born-Markov dynamics, which reduces the experimental difficulty. Another finding is that $\mathcal{E}_\text{B}(t)$ monotonically increases with time when $D$ increases from zero to one. Remarkably, the steady-state energy $\mathcal{E}_\text{B,ss}$ when $D=1$ is about 34 times greater than that when $D = 0$. This enhancement can be further increased by increasing the driving amplitude $\Omega$; see Appendix \ref{Appendix:driving strength}. It manifests the powerful charging performance enhanced by the nonreciprocity in our QB. Figure~\ref{energy}(b) reveals a similar increasing behavior for the charging efficiency $\eta(t)$ with $D$. The crossover to $\eta(t)> 1$, where $\mathcal{E}_\text{B}(t)$ exceeds $\mathcal{E}_\text{C}(t)$, marks a breakdown of reciprocity. This is a manifestation of a chirality-directed current that unidirectionally channels the energy from the charger to the QB. It underscores another dominant role played by the chirality in our QB. Figure~\ref{energy}(c) shows that the ergotropy $\mathcal{W}(t)$ evolves to a stable value that increases with $D$. A chirality enhanced gain as high as 55 is achieved over the achiral case. The work-extraction efficiency $R(t)$ in Fig.~\ref{energy}(d) saturates to 0.99. It verifies that almost all the QB energy can be converted to useful work in the maximally nonreciprocal case. The numerical results in Fig.~\ref{energy} agree with the steady-state solutions in Eqs. \eqref{Ecinfty}--\eqref{winfty}, see the cyan solid lines. The detailed dynamical solution is given in Appendix \ref{Dynamical solution}.

\begin{figure}
\includegraphics[width=0.98\columnwidth]{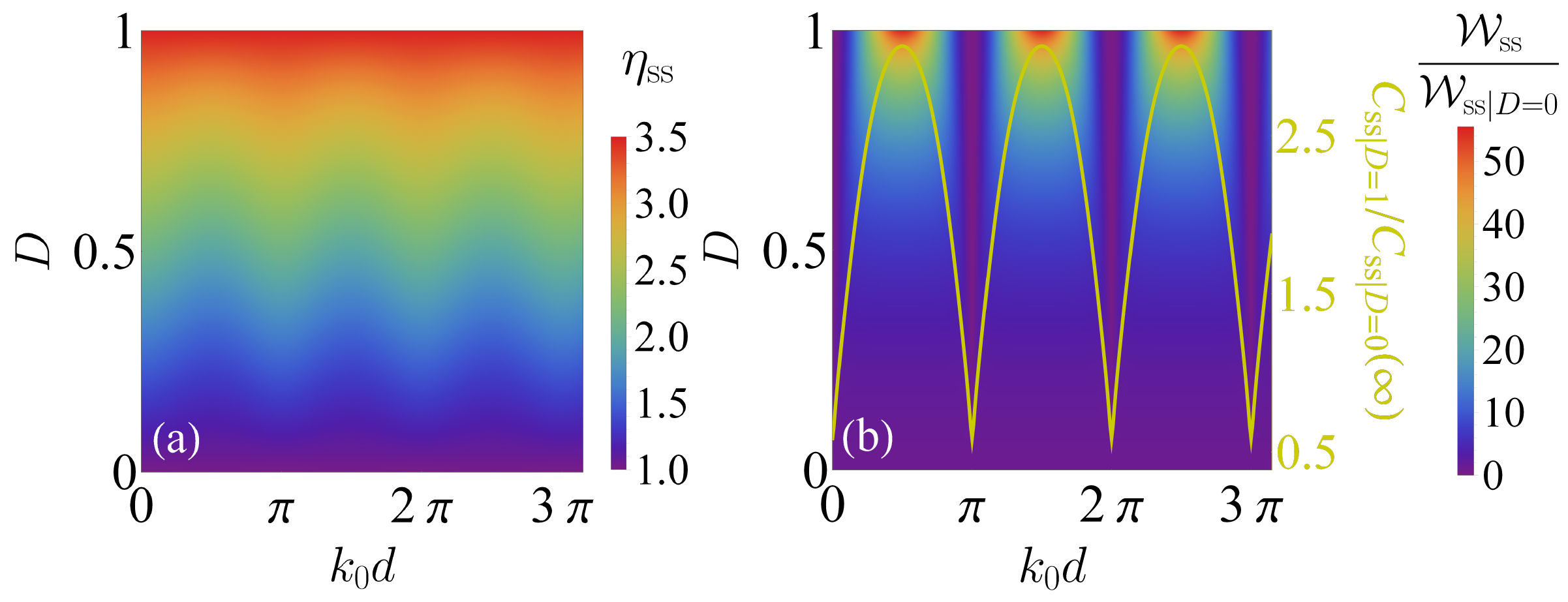}
\caption{(a) Charging efficiency $\eta_\text{ss}$ and (b) ratio of steady-state ergotropy $\mathcal{W}_\text{ss}/\mathcal{W}_{\text{ss}|D=0}$ as a function of the nonreciprocity parameter $D$ and the QB-charger distance $d$ when $\Gamma_L$ varies from $0$ to $\Gamma_R$. The corresponding quantum coherence $C_{\text{ss}|D=1}/C_{\text{ss}|D=0}$ is denoted by the yellow solid lines in (b). The parameters are the same as Fig. \ref{energy}.}\label{energyDkd}
\end{figure}

To elucidate the tolerance of the charging efficiency to the QB-charger distance $d$, we plot in Fig.~\ref{energyDkd}(a) the remote-charging efficiency $\eta_\text{ss}$ in the steady state as a function of $D$ and $d$. An observation is that $\eta_\text{ss}$ increases with $D$ while remaining almost independent of $d$ for large $D$. Thus, the remote-charging efficiency is governed by the nonreciprocity and the coherent interference modulated by the phase $k_{0}d$ has a negligible influence on the steady-state energy. As far as the charging is concerned, our QB works for an arbitrary QB-charger distance.
This enables robust wirelesslike remote charging of the QB over a broad range of charger-QB distances. Figure~\ref{energyDkd}(b) presents the chirality enhanced gain of the steady-state ergotropy with respect to the one in the achiral case, i.e., $\mathcal{W}_\text{ss}/\mathcal{W}_{\text{ss}|D=0}$, with the change of $D$ and $d$. It exhibits a periodic dependence on $d$. Its maxima occur at $d=(n+1/2)\pi/k_0$ and its minima are at $d=n\pi/k_{0}$, where $n$ is an integer. When $D=1$, $\mathcal{W}_\text{ss}/\mathcal{W}_{\text{ss}|D=0}$ reaches 55 at the constructive-interference distances $d=(n+1/2)\pi/k_0$ and tends to 1 at the destructive-interference distances $d=n\pi/k_0$. We also plot the ratio $C_{\text{ss}|D=1}/C_{\text{ss}|D=0}$ between the relative-entropy coherence of $D=1$ and that of $D=0$. It is interesting to find that $\mathcal{W}_\text{ss}/\mathcal{W}_{\text{ss}|D=0}$ is in synchronization with $C_{\text{ss}|D=1}/C_{\text{ss}|D=0}$. The maxima of the chirality-enhanced ergotropy exactly correspond to those of the coherence. In the present Gaussian model, the covariance matrix of the QB remains isotropic, and the relative-entropy coherence and the ergotropy are both determined by the same coherent displacement of the battery mode, see Appendix \ref{Appendix:Energies and ergotropy}. Therefore, their synchronized behavior is not sufficient to prove that quantum coherence drives the enhancement. This enhancement microscopically stems from a synergy between the nonreciprocity and coherent interference induced by the chirality. The nonreciprocity provides a directionality to the energy flow, whereas the coherent interference with the relative phase $k_0d$ accumulated via the waveguide propagation between the charger and the QB governs the battery amplitude and the ergotropy of the QB. For the remote-charging efficiency, the nonreciprocity acts primarily as a switch and amplifier of unidirectional energy flow. For the ergotropy, however, the nonreciprocity must cooperate with the spatial interference to determine the amount of extractable work from the QB. This offers a theoretical insight for the design of high-performance QB based on chiral systems.

\begin{figure}
\includegraphics[width=0.98\columnwidth]{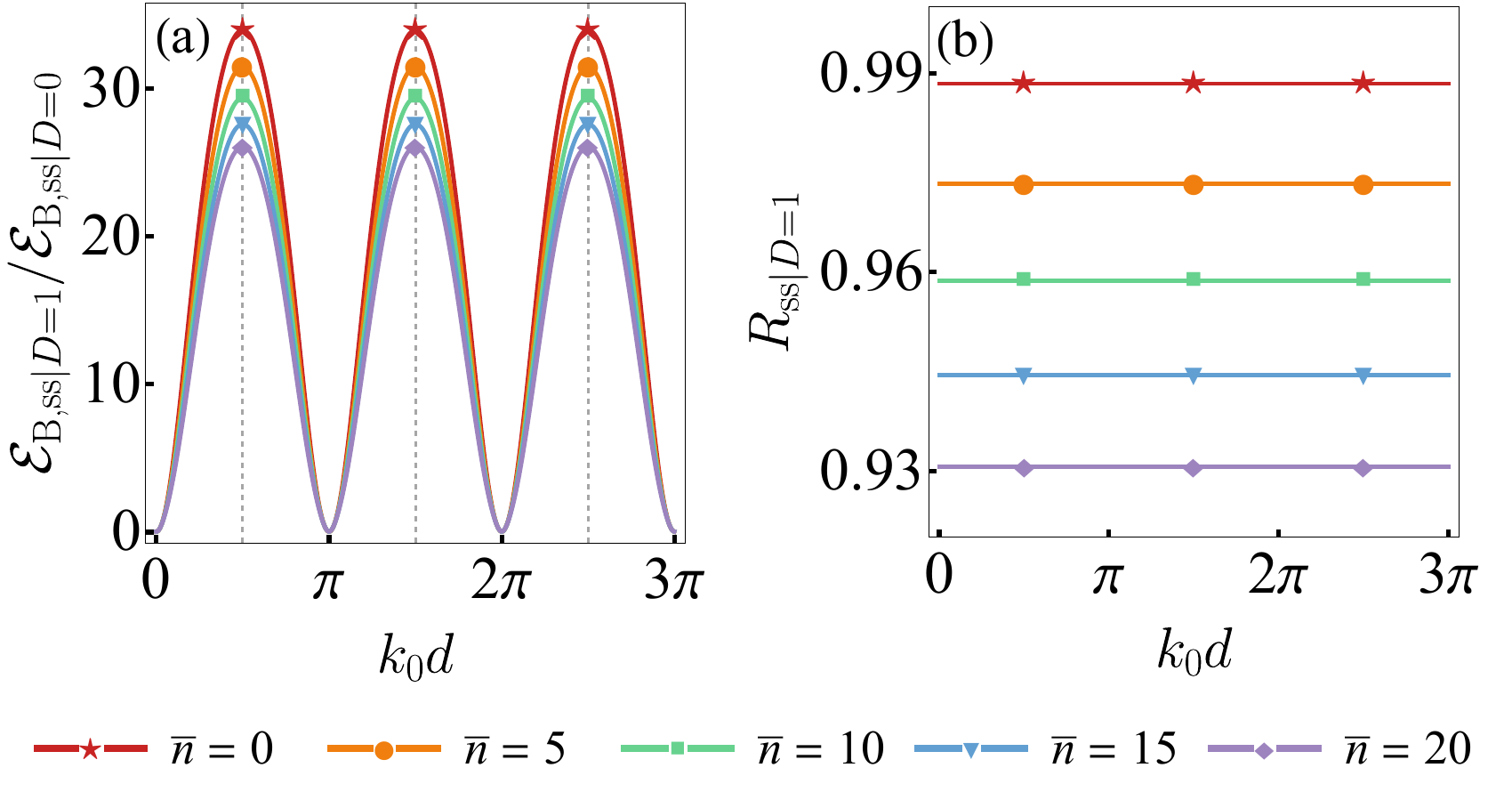}
\caption{(a) Chirality enhanced ratio of the steady-state QB energy $\mathcal{E}_{\text{B,ss}|D=1}/\mathcal{E}_{\text{B,ss}|D=0}$ and (b) $R_{\text{ss}|D=1}$ as a function of the QB-charger distance $d$ for different values of $\bar{n}$. Other parameters are the same as Fig. \ref{energy}.}\label{temperature}
\end{figure}

We now examine the thermal effects on the QB performance. Figure~\ref{temperature}(a) shows $\mathcal{E}_{\text{B,ss}|D=1}/\mathcal{E}_{\text{B,ss}|D=0}$ as a function of the QB-charger distance $d$ for different values of $\bar{n}$. The enhancement ratio is highest at zero temperature and decreases with increasing $\bar{n}$. Its periodic dependence on $d$ remains unchanged, reflecting the phase-sensitive interference associated with the propagation phase $k_0d$ between the charger and the QB. The steady-state efficiency as a function of $d$ for various $\bar{n}$ in Fig.~\ref{temperature}(b) shows $R_{\text{ss}|D=1}$ decreases as $\bar{n}$ increases. This trend follows from Eqs. \eqref{Ebinfty} and \eqref{winfty}: the ergotropy is independent of temperature, whereas the stored energy $\mathcal{E}_{\text{B,ss}}$ grows with temperature, thereby reducing $R_{\text{ss}|D=1}$ for higher $\bar{n}$. However, $R_{\text{ss}|D=1}$ remains constant with the QB-charger distance, demonstrating that the remote-charging capability of our QB is not degraded by temperature.

\begin{figure}[tbp]
\includegraphics[width=0.9\columnwidth]{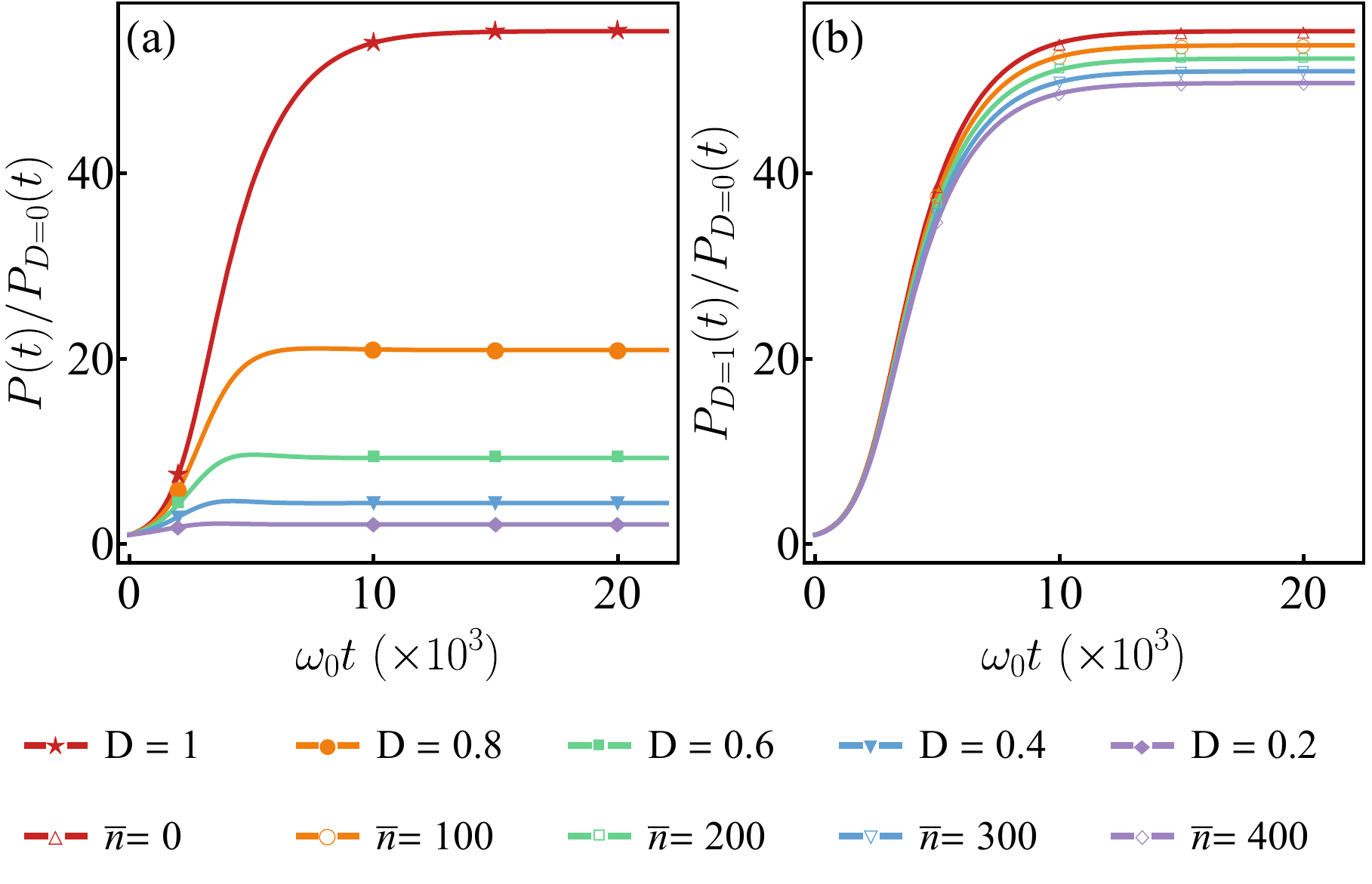}
\caption{Evolution of the charging power ratio in different (a) $D$ and (b) thermal excitation number $\bar{n}$ when $k_{0}d=\pi/2$. Other parameters are the same as Fig. \ref{energy}.}\label{chiralitypt}
\end{figure}

We now examine the charging power $P(t)=[\mathcal{E}_{\text{B}}(t)-\mathcal{E}_{\text{B}}(0)]/t$. As shown in Fig.~\ref{chiralitypt}(a), the chirality enhanced power ratio $P(t)/P_{D=0}(t)$ over the achiral case increases monotonically to a stable value. Large $D$ consistently accelerates the charging process and elevates the saturation threshold. Specifically, operating at $D=1$ yields a steady-state power enhancement factor of approximately $55$ relative to the achiral baseline ($D=0$). This factor matches the enhancement factor of ergotropy. To assess the thermal stability of this chiral advantage, we plot in Fig.~\ref{chiralitypt}(b) the time evolution of $P_{D=1}(t)/P_{D=0}(t)$ for different values of the thermal excitation number $\bar{n}$. The profile of monotonic-growth behavior remains invariant, with all curves converging to a substantially enhanced plateau. Increasing $\bar{n}$ causes only a marginal, slow decline in the power ratio. This weak dependence on the thermal excitation number shows that the chirality-driven power gain is highly robust against thermal excitations.

\begin{figure}[tbp]
\includegraphics[width=0.95\columnwidth]{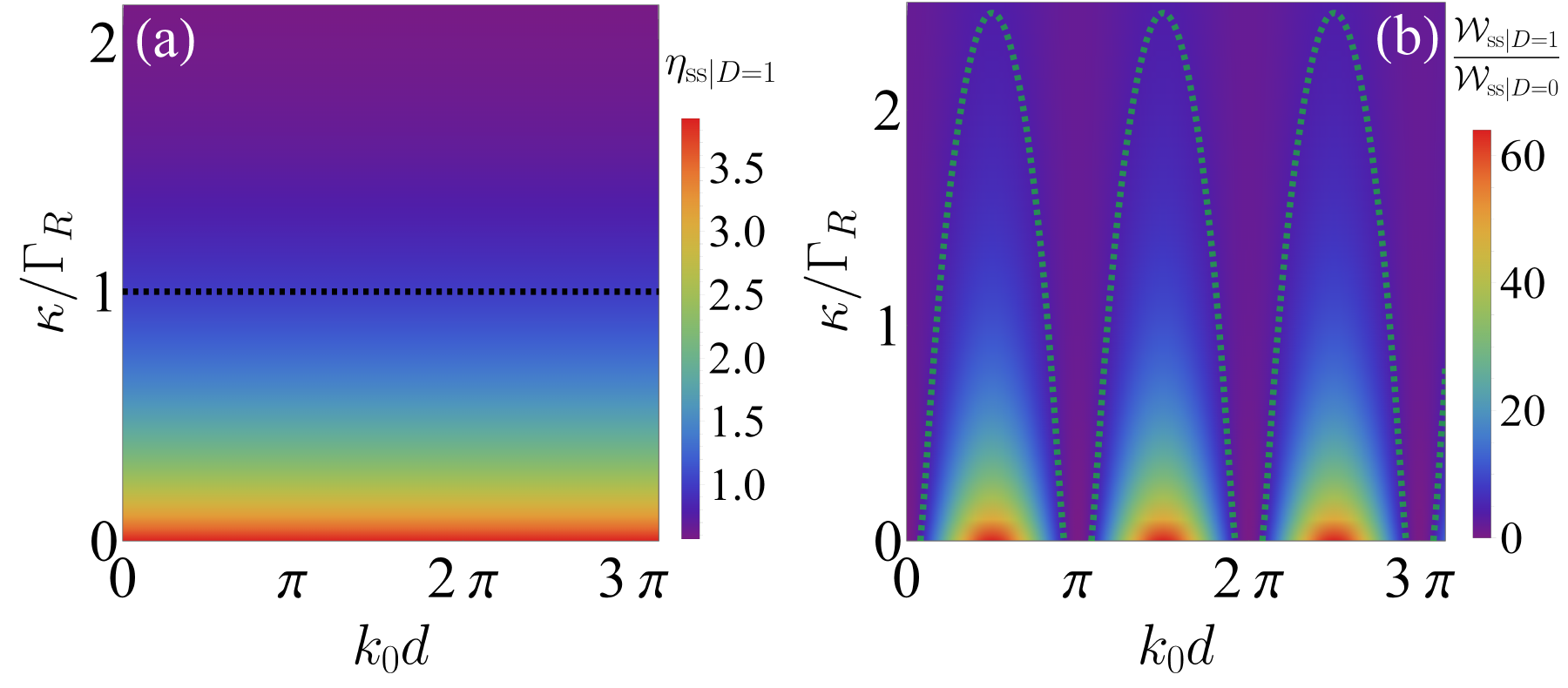}
\caption{(a) Charging efficiency $\eta_{\text{ss}|D=1}$ and (b) ergotropy ratio $\mathcal{W}_{\text{ss}|D=1}/\mathcal{W}_{\text{ss}|D=0}$ in the $(k_{0}d, \kappa/\Gamma_{R})$ plane. The black dashed line in (a) marks $\eta_{\text{ss}|D=1}=1$. The green dashed line in (b) indicates $\mathcal{W}_{\text{ss}|D=1}/\mathcal{W}_{\text{ss}|D=0}=4$. Other parameters are the same as Fig. \ref{energy}.}\label{chiralityhs}
\end{figure}

We have used $\kappa/2\pi = 1$~MHz in the numerical calculation, which corresponds to $\kappa/\Gamma_R = 0.05$ and is well within the practical operating regime of state-of-the-art magnon experiments \cite{PhysRevLett.128.047701}. To verify the tolerance of our scheme to larger dissipation rates, we plot the steady-state charging efficiency $\eta_{\text{ss}|D=1}$ and the ergotropy ratio $\mathcal{W}_{\text{ss}|D=1}/\mathcal{W}_{\text{ss}|D=0}$ as a function of distance $k_{0}d$ and $\kappa/\Gamma_R$ in Figs. \ref{chiralityhs}(a) and \ref{chiralityhs}(b), respectively. Figure~\ref{chiralityhs}(a) shows that $\eta_{\text{ss}|D=1}$ decreases with increasing $\kappa/\Gamma_{R}$ but shows little dependence on $k_{0}d$. 
The black dashed line marks $\eta_{\text{ss}|D=1}=1$. For $\kappa/\Gamma_{R} < 1$, we find $\eta_{\text{ss}|D=1} > 1$, i.e., the QB stores more energy than the charger. Even at $\kappa/2\pi=3$ MHz, $\eta_{\text{ss}|D=1} \approx 3$ remains. This result suggests that $\kappa/\Gamma_R \lesssim 1$ is a practically reasonable operating regime to maintain the chirality-induced remote charging. It confirms that our QB exhibits robustness across a wide range of dissipation rates. Figure~\ref{chiralityhs}(b) shows that our system maintains an ergotropy enhancement far beyond the fourfold benchmark in Ref.~\cite{PhysRevLett.132.210402} over a wide parameter regime. For $\kappa/\Gamma_R \lesssim 0.1$, $\mathcal{W}_{\text{ss}|D=1}/\mathcal{W}_{\text{ss}|D=0}$ remains in the range of 48--64, where 64 corresponds to the dissipation-free case. Therefore, although gradually vanishing with increasing $\kappa/\Gamma_R$, the chirality enhanced advantages for $\eta_{\text{ss}|D=1}$ and $\mathcal{W}_{\text{ss}|D=1}/\mathcal{W}_{\text{ss}|D=0}$ remain present in an extremely wide range of the dissipation rate of the magnons. As a final remark, the enhanced performance of our QB is insensitive to the imperfection of the positions of the two magnons, see Appendix \ref{impftc}.

\begin{figure}[tbp]
\includegraphics[width=0.9\columnwidth]{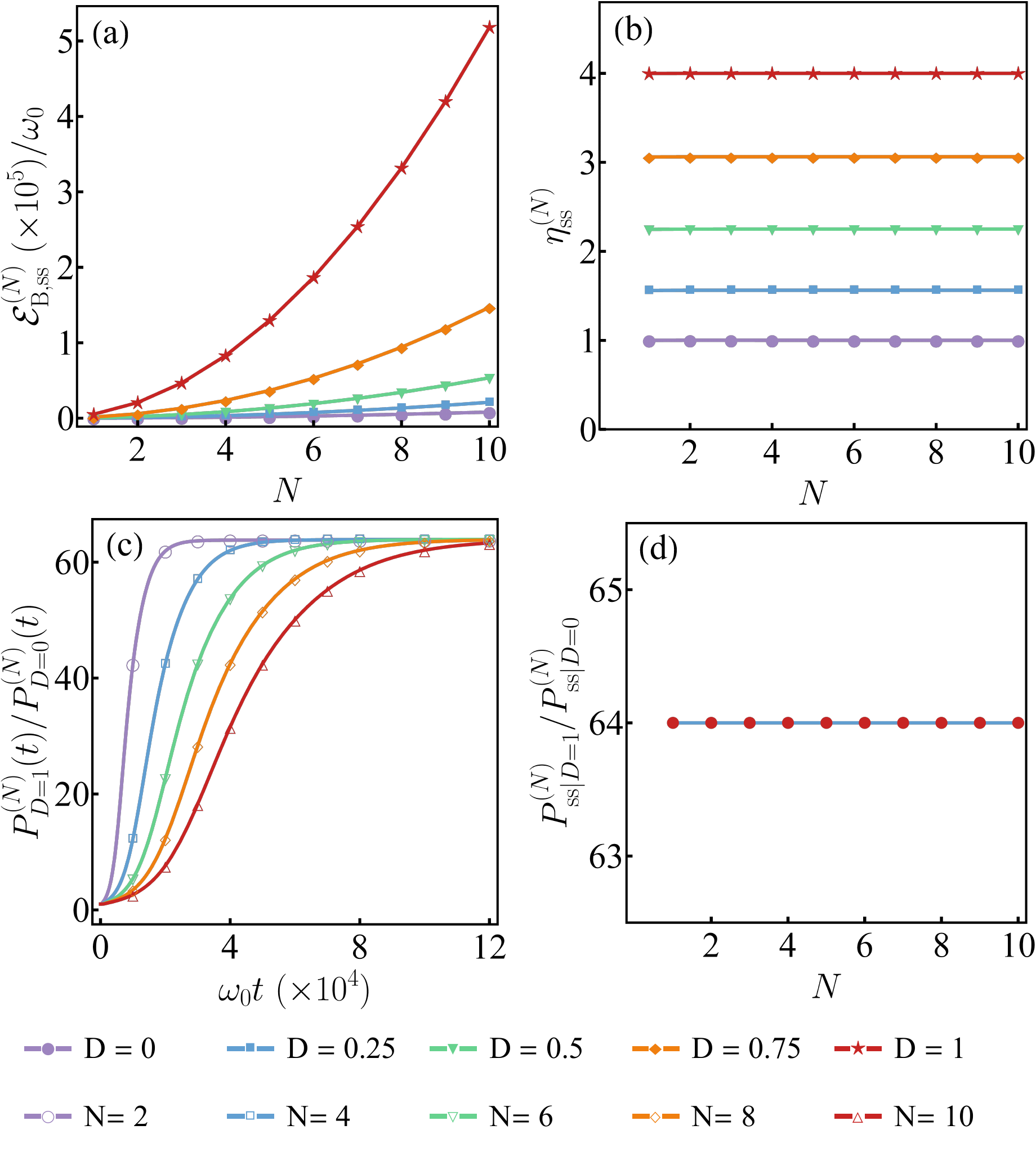}
\caption{(a) $\mathcal{E}_{\text{B,ss}}^{(N)}$ and (b) $\eta_{\text{ss}}^{(N)}$ as a function of $N$ in different $D$. (c) Evolution of $P_{D=1}^{(N)}(t)/P_{D=0}^{(N)}(t)$ in different $N$. (d) $P_{\text{ss}|D=1}^{(N)}/P_{\text{ss}|D=0}^{(N)}$ as a function of $N$. $k_{0}d=\pi/2$, $\kappa=0$, $g_{\lambda_1}(\omega_0)=g_{\lambda}(\omega_0)/\sqrt{N}$, $g_{\lambda_2}(\omega_0)=g_{\lambda}(\omega_0)/N$, and other parameters are the same as Fig. \ref{energy}.}\label{NN}
\end{figure}

\section{Scalability to multi-magnon QB}\label{scalbst}
With the scalable and tunable framework of waveguide magnonics system, our scheme is extendable to multi-magnon scenarios. We extend our QB to contain $N$ identical magnons. Since the submillimeter magnon spheres are significantly smaller than the centimeter-scale photon wavelength of the waveguide, we can approximately treat the magnons as point particles residing in the same spatial position \cite{PhysRevLett.124.107202}. The coupling strength between the charger magnon and the EMF is $g_{\lambda_1}$, while that between the $N$ QB magnons and the EMF is $g_{\lambda_2}$. By introducing the collective bosonic operator $\hat{\mathcal{B}}=N^{-1/2}\sum_{j}\hat{m}_{j}$ satisfying the commutation relation $[\hat{\mathcal{B}},\hat{\mathcal{B}}^{\dagger}]=1$ \cite{PhysRevLett.132.153602,doi:10.1126/science.adr8187}, we derive the master equation for the $N+1$-magnon model as
\begin{eqnarray}
&&\dot{\rho}(t)=-i[\tilde{H}_\text{m}(t)+\sum_{\lambda=L,R}\tilde{H}_{\lambda},\rho (t)]+\big\{\sum_{\lambda=L,R}\check{\mathcal{L}}_{\tilde{M}_{\lambda}} \nonumber\\
&&~~~~~~~~~~+\kappa\sum_{j=1}^{N+1}[(\bar{n}+1)\check{\mathcal{L}}_{\hat{m}_{j}} +\bar{n}\check{\mathcal{L}}_{\hat{m}^{\dagger }_{j}}]\}\rho(t),
\end{eqnarray}
where $\tilde{H}_{\text{m}}(t)=\omega_0\sum_{j}^{N+1}(\hat{m}_{j}^{\dagger}\hat{m}_{j}+\frac{1}{2})+\Omega(\hat{m}_{1}^{\dagger}e^{-i\omega_d t}+\text{H.c.})$, $\tilde{H}_{L}=-\frac{ig_{L_{1}}(\omega_{0})g_{L_{2}}(\omega_{0})}{2}\sqrt{N}(\hat{m}_{1}^{\dagger}\hat{\mathcal{B}}e^{ik_{0}d}-\text{H.c.})$, and $\tilde{H}_{R}=-\frac{ig_{R_{1}}(\omega_{0})g_{R_{2}}(\omega_{0})}{2}\sqrt{N}(\hat{m}_{1}\hat{\mathcal{B}}^{\dagger}e^{ik_{0}d}-\text{H.c.})$. The operators $\tilde{M}_L=g_{L_{1}}(\omega_{0})\hat{m}_{1}+g_{L_{2}}(\omega_{0})\sqrt{N}\hat{\mathcal{B}}e^{ik_{0}d}$ and $\tilde{M}_R=g_{R_{1}}(\omega_{0})\hat{m}_{1}+g_{R_{2}}(\omega_{0})\sqrt{N}\hat{\mathcal{B}}e^{-ik_{0}d}$ describe the collective dissipation induced by the left- and right-propagating EMFs in the waveguide. 

To evaluate the scaling behavior under the coupling parametrization specified above, we consider the ideal dissipation-free limit ($\kappa=0$). At resonance, $\omega_d=\omega_0$, and under the coupling parametrization
$g_{\lambda_1}(\omega_0)=g_\lambda(\omega_0)/\sqrt{N}$ and $g_{\lambda_2}(\omega_0)=g_\lambda(\omega_0)/N$, the equations of motion for the first moments reduce to
\begin{eqnarray}
\frac{d\langle\hat m_1\rangle}{dt}&=&
-\frac{\Gamma_L+\Gamma_R}{2N}\langle\hat m_1\rangle
-\frac{\Gamma_L}{N}e^{ik_0d}\langle\hat{\mathcal B}\rangle
-i\Omega ,\\
\frac{d\langle\hat{\mathcal B}\rangle}{dt}&=&
-\frac{\Gamma_L+\Gamma_R}{2N}\langle\hat{\mathcal B}\rangle
-\frac{\Gamma_R}{N}e^{ik_0d}\langle\hat m_1\rangle .
\end{eqnarray}
The corresponding steady-state solutions are
\begin{eqnarray}
\langle\hat m_1\rangle_{\rm ss}&=&-\frac{2i\Omega N(\Gamma_L+\Gamma_R)}
{(\Gamma_L+\Gamma_R)^2-4\Gamma_L\Gamma_Re^{2ik_0d}},\\
\langle\hat{\mathcal B}\rangle_{\rm ss}&=&\frac{4i\Omega N\Gamma_Re^{ik_0d}}
{(\Gamma_L+\Gamma_R)^2-4\Gamma_L\Gamma_Re^{2ik_0d}}.
\label{eq:bss_multi}
\end{eqnarray}
The steady-state battery energy and ergotropy are expressed in terms of the corresponding single-magnon results as
\begin{eqnarray}
\mathcal{E}_{\rm B,ss}^{(N)}&=&\frac{N\omega_0}{2}+N^{2}(\mathcal{E}_{\rm B,ss}-\frac{\omega_0}{2}),
\label{eq:EB_multi_ss}\\
\mathcal{W}_{\rm ss}^{(N)}&=&N^{2}\mathcal{W}_{\rm ss}.
\label{eq:W_multi_ss}
\end{eqnarray}

Figures~\ref{NN}(a) and \ref{NN}(b) show the steady-state QB energy $\mathcal{E}_\text{B,ss}^{(N)}$ and remote-charging efficiency $\eta_\text{ss}^{(N)}$ as functions of the magnon number $N$ for different nonreciprocity parameters $D$, respectively. The numerical results for $\mathcal{E}_\text{B,ss}^{(N)}$ agree well with the analytical expression in Eq.~(\ref{eq:EB_multi_ss}), consistent with the quadratic scaling of the charging-induced energy. The corresponding ergotropy follows the same $N^2$ scaling according to Eq.~(\ref{eq:W_multi_ss}), while the steady-state remote-charging efficiency $\eta_\text{ss}^{(N)}$ remains independent of $N$.
Figure~\ref{NN}(c) illustrates the transient dynamics of the chiral advantage through the power ratio $P_{D=1}^{(N)}(t)/P_{D=0}^{(N)}(t)$. The power ratios for all $N$ converge to the same steady-state plateau, while increasing $N$ slows the transient evolution. Figure~\ref{NN}(d) shows that the steady-state power ratio $P_{\text{ss}|D=1}^{(N)}/P_{\text{ss}|D=0}^{(N)}$ is independent of $N$ and remains at 64, in agreement with the analytical steady-state results. These findings demonstrate that our multi-magnon QBs quadratically enhance the energy capacity without degrading long-term charging efficiency. Therefore, the coupled magnon-waveguide system provides a useful platform for the practical realization of large-scale QBs.

\section{Discussion and conclusion}\label{sec:conclusion}
In contrast to the nonreciprocal QB that requires the balance between charger-QB interactions and local dissipations \cite{PhysRevLett.132.210402}, the chirality arises from the spin-momentum locking of the light-magnon interactions \cite{WOS000503012600001,Qian2025}. It grants our QB unique capabilities, including wirelesslike remote charging, robustness against local dissipations, and operation without fine-tuning of charger-QB coupling. The synergy between chirality-induced nonreciprocal energy transfer and phase-sensitive coherent interference leads to a remarkable enhancement of ergotropy. These features make chiral magnon-waveguide coupling a promising architecture for high-performance QBs. Current experimental advances in chiral waveguide QED provide support for our QB scheme \cite{PhysRevLett.134.173601,PhysRevLett.127.233601}. Based on the cavity magnonics and waveguide magnonics system \cite{PhysRevLett.124.107202,PhysRevB.107.064401,PhysRevB.101.094414}, a series of integrated platforms have ensured the feasibility of our proposal, such as a ferrimagnetic insulator with high spin density and low damping rates, enabling strong and ultrastrong coupling between magnetostatic modes in YIG spheres and microwave cavity modes \cite{PhysRevLett.113.156401,PhysRevLett.120.057202,PhysRevLett.123.127202,PhysRevLett.124.053602,PhysRevLett.128.047701,PhysRevLett.132.206902,PhysRevLett.134.196904,PhysRevLett.135.186704}. Experimental observation has confirmed the chiral magnons in altermagnetic systems \cite{PhysRevLett.135.126702,PhysRevLett.135.186703}. In addition to the chiral coupling of emitters to light fields \cite{coles2016chirality,PhysRevLett.126.233602}, the strong magnon-photon coupling in a microwave waveguide \cite{DEMIDOV20171}, the negative refractive index effect in photon-magnon system \cite{PhysRevLett.135.116905}, and the Kittel mode based on ring cavities or waveguides \cite{PhysRevLett.124.107202,PhysRevApplied.19.014030,PhysRevB.106.104432,PhysRevB.101.094414} also set a solid foundation for the realizability of our scheme. 

In conclusion, we have proposed a continuous-variable chiral QB scheme based on the waveguide magnonics system, where the charger and the QB are formed by two distant YIG spheres embedded in the waveguide. The unique chiral interactions between the magnons of the YIG spheres and the EMFs in the waveguide induce the nonreciprocity of the energy flow from the charger to the QB and the coherent interference between them without the inversion symmetry. Due to the interplay of these two effects, our QB possesses the capabilities of perfect remote-charging and robustness to the magnon dissipation. The ergotropy of our QB has been found to exhibit a chirality-enhanced gain factor of 55, approaching its theoretical limit of 64, within the experimentally accessible parameter regime. The proposed chiral magnonic architecture thus provides a route toward directional and wirelesslike energy transfer in continuous-variable QBs. For the vacuum initial state considered here, no charger-QB entanglement is generated throughout the charging dynamics. Because the state remains Gaussian, the charging process is classically representable. Accordingly, under the operating conditions considered here, the enhanced charging performance can be achieved without nonclassical-state preparation, which may facilitate the experimental implementation of the proposed scheme. With the scalable and tunable framework of waveguide magnonics, our scheme can be extended to multi-magnon configurations. The scaling behavior of such extensions depends on the microscopic coupling parametrization, as illustrated by the conditional $N^2$ scaling discussed above.

\section*{Acknowledgments}
The work is supported by the Quantum Science and Technology-National Science and Technology Major Project (Grants No. 2023ZD0300404 and No. 2023ZD0300904), the National Natural Science Foundation of China (Grants No. 12275109, No. 92576202, No. 12274422, No. U25A20196, and No. 12247101), the Natural Science Foundation of Hubei Province (2022CFA013), and the Natural Science Foundation of Gansu Province (Grants No. 26RCKA011 and No. 25JRRA799).
\appendix

\section{Master equation}\label{Appendix:Master equation}
The total system consists of the two magnon modes under classical driving, the EMFs in the waveguide, and two environments. The free Hamiltonian is
\begin{eqnarray}
\hat{H}_{0} &=&\sum_{j=1,2}[\omega _{0}(\hat{m}_{j}^{\dagger }\hat{m}_{j}+\frac{1}{2})+\int \omega \hat{c}_{j}^{\dagger }(\omega )\hat{c}_{j}(\omega)d\omega ]\nonumber\\
&&+\sum_{\lambda =L,R}\int \omega \hat{b}_{\lambda }^{\dagger}(\omega )\hat{b}_{\lambda }(\omega )d\omega,
\end{eqnarray}%
where $\hat{c}_{j}(\omega )$ and $\hat{b}_{\lambda}(\omega )$ are the annihilation operators of the $j$th environment and the $\lambda$-propagating EMF. Their interaction Hamiltonian in the interaction picture with respect to $\hat{H}_0$ is $\hat{H}^{I}(t)=\hat{H}_{d}(t)+\hat{H}'(t)$, where
\begin{eqnarray}
\hat{H}'(t)&=&\hat{H}_\text{F}(t)+\hat{H}_\text{E}(t),\label{feint}\\
\hat{H}_{d}(t) &=&\Omega [\hat{m}_{1}^{\dagger }e^{-i(\omega_{d}-\omega_0)t}+\text{H.c.}],  \label{EqA2} \\
\hat{H}_\text{F}(t) &=&\sum_{\substack{ \lambda =L,R  \\ j=1,2}}\int \frac{id\omega }{\sqrt{2\pi }}[g_{\lambda j}(\omega )\hat{b}_{\lambda }^{\dagger}(\omega )\hat{m}_{j}e^{i(\omega -\omega_{0})t-\frac{i\omega z_{j}}{v_{\lambda }}} \nonumber \\
&&-\text{H.c.}], \label{incdd}\\
\hat{H}_\text{E}(t) &=&\sum_{j=1,2}\int \frac{id\omega }{\sqrt{2\pi }}[\chi(\omega )\hat{c}_{j}^{\dagger }(\omega )\hat{m}_{j}e^{i(\omega -\omega_{0})t}-\text{H.c.}].~~~~~\label{indcccs}
\end{eqnarray}
Initially, the EMFs are in the vacuum state and the environments are in the thermal state at a finite temperature, i.e., $\rho_\text{F}=\prod_{\lambda,\omega}|0_\lambda(\omega)\rangle\langle0_\lambda(\omega)|$ and $\rho_\text{E}=\prod_j\exp[-\int\omega \hat{c}_j^\dag(\omega)\hat{c}_j(\omega)/(k_BT) d\omega]/Z$. Therefore, we have the expectation values of the operators as 
\begin{eqnarray}
&&\text{Tr}_\text{F}[\hat{b}_\lambda(\omega)\rho_\text{F}]=\text{Tr}_\text{E}[\hat{c}_j(\omega)\rho_\text{E}]=0,\label{fstmm}\\
&&\text{Tr}_\text{F}[\hat{b}_{\lambda }^{\dagger }(\omega )\hat{b}_{\lambda ^{\prime }}(\omega ^{\prime})\rho_\text{F}] =\text{Tr}_\text{F}[\hat{b}_{\lambda }(\omega )\hat{b}_{\lambda ^{\prime }}(\omega^{\prime})\rho_\text{F}] =0,  ~~~~\\
&&\text{Tr}_\text{F}[ \hat{b}_{\lambda }(\omega )\hat{b}_{\lambda ^{\prime }}^{\dagger}(\omega ^{\prime })\rho_\text{F}] =\delta (\omega -\omega ^{\prime })\delta_{\lambda ,\lambda ^{\prime }}, \\
&&\text{Tr}_\text{E}[\hat{c}_{j}^{\dagger }(\omega )\hat{c}^{\dagger }_{j^{\prime}}(\omega ^{\prime })\rho_\text{E}] =\text{Tr}_\text{E}[ \hat{c}_{j}(\omega )\hat{c}_{j^{\prime }}(\omega ^{\prime})\rho_\text{E}]  =0,   \\
&&\text{Tr}_\text{E}[\hat{c}_{j}(\omega )\hat{c}_{j^{\prime }}^{\dagger }(\omega^{\prime })\rho_\text{E}] =(\bar{n}_{\omega}+1)\delta(\omega -\omega ^{\prime })\delta_{j,j^{\prime }},  \\
&&\text{Tr}_\text{E}[\hat{c}_{j}^{\dagger }(\omega )\hat{c}_{j^{\prime }}(\omega^{\prime })\rho_\text{E}] =\bar{n}_{\omega}\delta (\omega -\omega ^{\prime })\delta_{j,j^{\prime }},
\end{eqnarray}
where $\bar{n}_\omega=[e^{\omega/(k_BT)}-1]^{-1}$. The Liouville equation is $\dot{\tilde\rho}_\text{tot}(t)=-i[\hat{H}^{I}(t),{\tilde\rho}_\text{tot}(t)]$ under the initial condition ${\tilde\rho}_\text{tot}(0)=\rho_\text{F}\rho_\text{E}\rho(0)$. Substituting its solution ${\tilde\rho}_\text{tot}(t)={\tilde\rho}_\text{tot}(0)-i\int_0^td\tau[\hat{H}^{I}(\tau),{\tilde\rho}_\text{tot}(\tau)]$ back into it and tracing the EMFs and the environments, we have 
\begin{eqnarray}
\dot{\tilde\rho}(t)&=&-i[\hat{H}_d(t),{\tilde\rho}(t)]-\text{Tr}_\text{F,E}\int_0^td\tau  \nonumber\\
&&~~~\times[\hat{H}'(t),[\hat{H}_d(\tau)+\hat{H}'(\tau),{\tilde\rho}_\text{tot}(\tau)]],~~\label{dfded0}
\end{eqnarray}
where ${\tilde\rho}(t)=\text{Tr}_\text{F,E}[{\tilde\rho}_\text{tot}(t)]$. Under the Born approximation, ${\tilde\rho}_\text{tot}(\tau)$ factors as ${\tilde\rho}_\text{tot}(\tau)\simeq {\tilde \rho}(\tau)\rho_\text{F}(0)\rho_\text{E}(0)$. Furthermore, under the Markov approximation, we can replace ${\tilde\rho}(\tau)$ by ${\tilde\rho}(t)$. Equation \eqref{dfded0} is converted into \cite{Scully_Zubairy_1997}
\begin{eqnarray}
\dot{\tilde\rho}(t)&=&-i[\hat{H}_d(t),{\tilde\rho}(t)]-\text{Tr}_\text{F,E}\int_{0}^{t}d\tau\nonumber\\
&& \times\lbrack \hat{H}'(t),[\hat{H}'(\tau),{\tilde\rho}(t)\rho_\text{F}\rho_\text{E}]],\label{msct}
\end{eqnarray}
where Eq. \eqref{fstmm} has been used. 

Substituting Eq. \eqref{feint} into Eq. \eqref{msct}, we rewrite the last term of Eq. \eqref{msct} as $Q_1+Q_2+Q_3$, where 
\begin{eqnarray}
Q_{1/2} &=&-\text{Tr}_\text{F,E}\int_{0}^{t}d\tau  \lbrack H_\text{F/E}(t),[H_\text{F/E}(\tau ),\tilde{\rho}(t)\rho_\text{F}\rho_\text{E}]], ~~~~~~~\label{EqA7} \\
Q_{3} &=&-\text{Tr}_\text{F,E}\int_{0}^{t}d\tau \{\lbrack \hat{H}_\text{F}(t),[\hat{H}_\text{E}(\tau ),{\tilde\rho}(t)\rho_\text{F}\rho_\text{E}]]   \nonumber\\
&&+[\hat{H}_\text{E}(t),[\hat{H}_\text{F}(\tau ),{\tilde\rho}(t )\rho_\text{F}\rho_\text{E}]]\}. 
\end{eqnarray}
It is easy to verify $Q_3=0$ according to Eq. \eqref{fstmm}. The substitution of Eq. \eqref{incdd} into Eq. \eqref{EqA7} results in
\begin{widetext}
\begin{eqnarray}
Q_{1}
&=&\sum_{\substack{ \lambda =L,R \\ j,j'=1,2}}\int_{0}^{t}d\tau \int \frac{d\omega }{2\pi }\{g_{\lambda j'}(\omega )g_{\lambda j}^{\ast }(\omega )[\hat{m}_{j'}\tilde{\rho} (t)\hat{m}_{j}^{\dagger }-\hat{m}_{j}^{\dagger }\hat{m}_{j'}\tilde{\rho} (t)]e^{i\omega _{0}(t-\tau )}e^{-i\omega( \frac{z_{j'}-z_{j}}{v_{\lambda }}+t-\tau)} +\text{H.c.}\}.\label{mssat}
\end{eqnarray}
\end{widetext}
According to the Weisskopf-Wigner approximation, which is in the same accuracy as the Born-Markov approximation, $g_{\lambda l}(\omega)$ varies little around $ g_{\lambda l}(\omega_{0})$ for which the time integral in Eq. \eqref{mssat} is not negligible. We thus replace $g_{\lambda l}(\omega)$ by  $g_{\lambda l}(\omega_0)$ and the lower limit in the $\omega$ integration to $-\infty$ \cite{Scully_Zubairy_1997}. By using $\int \frac{d\omega }{2\pi}e^{\pm i\omega x}=\delta(x)$, Eq. \eqref{mssat} becomes 
\begin{eqnarray}
Q_{1}=\sum_{\substack{ \lambda =L,R  \\ j,j'=1,2}}\int_{0}^{t}d\tau\delta (\tau-\frac{
	z_{j'}-z_{j}}{v_{\lambda }}-t )g_{\lambda j}(\omega _{0})g_{\lambda j'}^{\ast }(\omega _{0})\nonumber\\
~~~~~~\times \{[\hat{m}_{j}\tilde{\rho} (t )\hat{m}_{j'}^{\dagger }-\tilde{\rho} (t )\hat{m}_{j'}^{\dagger }\hat{m}_{j}]e^{-i\omega _{0}(t-\tau)} +\text{H.c.}\}.~~~
\end{eqnarray}
For the $\text{TE}_{10}$ mode, the coupling strength becomes $g_{\lambda j}(\omega)=\sqrt{\frac{\gamma MV_{j}}{2\epsilon_{0}\omega ab}}[\frac{\pi}{a}\cos(\frac{\pi x_{j}}{a})-k_{\lambda}\sin(\frac{\pi x_{j}}{a})]$ \cite{PhysRevLett.124.107202,PhysRevB.101.094414}, where $a$ and $b$ are the lengths of the cross section of the waveguide, $\epsilon_{0}$ is the vacuum permittivity, and $k_\lambda=\omega/v_\lambda$ is the wave vector. We consider that the two YIG spheres have $x_1=x_2\equiv x_0$ and $V_1=V_2\equiv V$, which reduce $g_{\lambda1}(\omega_0)=g_{\lambda2}(\omega_0)\equiv g_{\lambda}(\omega_0)$. Defining $v_R=-v_L\equiv v$, $\omega_0/v\equiv k_0$, $d\equiv z_2-z_1$, and using \cite{10.1093/acprof:oso/9780198563617.001.0001}
\begin{equation}
\int_{x_1}^{x_2}f\left(x\right) \delta \left( x-x_{0}\right)dx =\left\{
\begin{array}{c}
f\left( x_{0}\right) ,\ \ \ x_{0}\in \left( x_1,x_2\right)  \\
\frac{f\left( x_{0}\right) }{2},\ \ \ x_{0}=x_1~\text{or}~x_2 \\
0,\text{ \ \ }x_{0}>x_2~\text{or}~x_{0}<x_1
\end{array}\right.,
\end{equation}
we derive
\begin{eqnarray}
Q_{1}&=&\sum_{\substack{ \lambda =L,R  \\ j=1,2}}\frac{\Gamma_\lambda}{2}[2\hat{m}_{j}\tilde{\rho} (t)\hat{m}_{j}^{\dagger }-\hat{m}_{j}^{\dagger }\hat{m}_{j}\tilde{\rho} (t)-\tilde{\rho} (t)\hat{m}_{j}^{\dagger }\hat{m}_{j}] \nonumber\\
&&+\Gamma_R\{
[ \hat{m}_{2}\tilde{\rho} (t)\hat{m}_{1}^{\dagger }-\tilde{\rho} (t)\hat{m}_{1}^{\dagger }\hat{m}_{2}]e^{-ik_{0}d}+\text{H.c.}\} ~~~\nonumber\\
&&+\Gamma_L\{[ \hat{m}_{1}\tilde{\rho} (t)\hat{m}_{2}^{\dagger }-\tilde{\rho} (t)\hat{m}_{2}^{\dagger }\hat{m}_{1}]e^{-ik _{0}d}+\text{H.c.}\}~~~\nonumber\\
&=&-i[\sum_{\lambda=L,R}\hat{H}_{\lambda},\tilde{\rho}(t)]+\sum_{\lambda =L,R}\Gamma_{\lambda}\check{\mathcal{L}}_{\hat{M}_{\lambda}}\tilde{\rho}(t),
\end{eqnarray}
where $\Gamma_{\lambda}=g_{\lambda}(\omega_0)^{2}$ and $\check{\mathcal{L}}_{\hat{o}}\cdot = \hat{o}\cdot\hat{o}^\dagger - \frac{1}{2}\{\hat{o}^\dagger\hat{o}, \cdot\}$ is the Lindblad superoperator. $\hat{H}_{L}=-\frac{i\Gamma_{L}}{2}(\hat{m}_{1}^{\dagger}\hat{m}_{2}e^{ik_0d}-\text{H.c.})$ and $\hat{H}_{R}=-\frac{i\Gamma_{R}}{2}(\hat{m}_{2}^{\dagger}\hat{m}_{1}e^{ik_0d}-\text{H.c.})$ describe the coherent coupling between the magnons mediated by the left- and right-propagating EMFs, respectively. The operators $\hat{M}_L=\hat{m}_{1}+\hat{m}_{2}e^{ik_{0}d}$ and $\hat{M}_R=\hat{m}_{1}+\hat{m}_{2}e^{-ik_{0}d}$ describe the collective dissipations induced by the left- and right-propagating EMFs in the waveguide. 

In a similar procedure as above, we derive 
\begin{eqnarray}
Q_{2}=\sum_{j=1,2}\kappa\{(\bar{n}+1)\check{\mathcal{L}}_{\hat{m}_{j}}+\bar{n}\check{\mathcal{L}}_{\hat{m}^{\dagger }_{j}}\}{\tilde\rho}(t), 
\end{eqnarray}
with $\kappa\equiv|\chi(\omega_0)|^{2}$ and $\bar{n}\equiv\bar{n}_{\omega_0}$. Therefore, the master equation in the interaction picture becomes
\begin{eqnarray}
&&\dot{\tilde\rho}(t)=-i[\hat{H}_{d}(t)+\sum_{\lambda=L,R}\hat{H}_{\lambda},{\tilde\rho}(t)]+\big\{\sum_{\lambda =L,R}\Gamma_{\lambda}\check{\mathcal{L}}_{\hat{M}_{\lambda}}\nonumber \\
&&~~~~~~~~~~+\sum_{j=1,2}\kappa [(\bar{n}+1)\check{\mathcal{L}}_{\hat{m}_{j}}+\bar{n}\check{\mathcal{L}}_{\hat{m}^{\dagger }_{j}}]\}\tilde\rho(t),\label{appEq2}
\end{eqnarray}
Going back to the Schr\"{o}dinger picture, we obtain
\begin{eqnarray}
&&\dot{\rho}(t)=-i[\hat{H}_\text{m}(t)+\sum_{\lambda=L,R}\hat{H}_{\lambda},\rho (t)]+\big\{\sum_{\lambda=L,R}\Gamma_{\lambda}\check{\mathcal{L}}_{\hat{M}_{\lambda}}\nonumber  \nonumber\\
&&~~~~~~~~~~+\sum_{j=1,2}\kappa[(\bar{n}+1)\check{\mathcal{L}}_{\hat{m}_{j}} +\bar{n}\check{\mathcal{L}}_{\hat{m}^{\dagger }_{j}}] \}\rho(t),
\end{eqnarray}
where $\hat{H}_{\text{m}}(t)=\sum_{j=1,2}\omega_0(\hat{m}_{j}^{\dagger}\hat{m}_{j}+\tfrac{1}{2})+\Omega(\hat{m}_{1}^{\dagger}e^{-i\omega_d t}+\text{H.c.})$. 

\section{Dynamical solution}\label{Dynamical solution}

From the master Eq. \eqref{appEq2}, we obtain the closed equations of motion for the relevant operators as
\begin{eqnarray}
\frac{d\langle \hat{m}_{1}\rangle}{dt} &=&-\frac{\alpha}{2} \langle \hat{m}_{1}\rangle -\Gamma _{L}e^{ik_{0}d}\langle \hat{m}_{2}\rangle -i\Omega, \label{appm1}\\
\frac{d\langle \hat{m}_{1}^2\rangle}{dt} &=&-\alpha \langle \hat{m}_{1}^2\rangle -2\Gamma _{L}e^{ik_{0}d}\langle \hat{m}_{1}\hat{m}_{2}\rangle \nonumber\\
&&-2i\Omega \langle \hat{m}_{1}\rangle, \label{appm1p}\\
\frac{d\langle \hat{m}_{1}^{\dagger}\hat{m}_{1}\rangle}{dt} &=&-\alpha\langle \hat{m}_{1}^{\dagger}\hat{m}_{1}\rangle -(\Gamma_{L}e^{ik_{0}d}\langle \hat{m}_{1}^{\dagger}\hat{m}_{2}\rangle\nonumber \\&&+i\Omega\langle \hat{m}_{1}^{\dagger}\rangle +\text{c.c.})+\kappa \bar{n}, \label{apmm1}\\
\frac{d\langle \hat{m}_{2}\rangle}{dt} &=&-\frac{\alpha}{2} \langle \hat{m}_{2}\rangle -\Gamma _{R}e^{ik_{0}d}\langle \hat{m}_{1}\rangle, \label{appm2}\\
\frac{d\langle \hat{m}_{2}^2\rangle}{dt} &=&-\alpha \langle \hat{m}_{2}^2\rangle -2\Gamma _{R}e^{ik_{0}d}\langle \hat{m}_{1}\hat{m}_{2}\rangle, \label{appm2p}\\
\frac{d\langle \hat{m}_{2}^{\dagger}\hat{m}_{2}\rangle}{dt} &=&-\alpha\langle \hat{m}_{2}^{\dagger}\hat{m}_{2}\rangle -\Gamma_{R}(e^{ik_{0}d}\langle \hat{m}_{1}\hat{m}_{2}^{\dagger }\rangle\nonumber\\&&+\text{c.c.}) +\kappa \bar{n},\label{apmm2} \\
\frac{d\langle \hat{m}_{1}\hat{m}_{2}\rangle}{dt} &=&-\alpha \langle \hat{m}_{1}\hat{m}_{2}\rangle -\Gamma _{R}e^{ik_{0}d}\langle \hat{m}_{1}^2\rangle\nonumber\\&& -\Gamma _{L}e^{ik_{0}d}\langle \hat{m}_{2}^2\rangle-i\Omega \langle \hat{m}_{2}\rangle, \label{appm12}\\
\frac{d\langle \hat{m}_{1}^{\dagger}\hat{m}_{2}\rangle }{dt}&=&-\alpha\langle \hat{m}_{1}^{\dagger}\hat{m}_{2}\rangle -\Gamma_{R}e^{ik_{0}d}\langle \hat{m}_{1}^{\dagger}\hat{m}_{1}\rangle\nonumber\\
&&-\Gamma_{L}e^{-ik_{0}d}\langle \hat{m}_{2}^{\dagger}\hat{m}_{2}\rangle +i\Omega\langle \hat{m}_{2}\rangle, \label{apm1m2}
\end{eqnarray}
where $\alpha=\Gamma_{L}+\Gamma_{R}+\kappa$ and $\langle\cdot\rangle=\text{Tr}[\cdot\rho(t)]$. We consider that both the charger and the battery are initially prepared in the vacuum state $|0\rangle$. Consequently, the initial values of all the above moments are zero at $t=0$.

We apply the Laplace transform $\mathcal{L}(f)=\int_{0}^{\infty}f(t)e^{-st}dt$ to Eqs. \eqref{appm1} and \eqref{appm2} and obtain
\begin{eqnarray}
\mathcal{L}(\langle\hat{m}_{1}\rangle) &=&\frac{-i\Omega (s+\frac{\alpha}{2} )}{s(s-s_{+})(s-s_{-})}, \\
\mathcal{L}(\langle \hat{m}_{2}\rangle) &=&\frac{i\Omega \Gamma_{R}e^{ik_{0}d}}{s(s-s_{+})(s-s_{-})}.
\end{eqnarray}
where $s_{\pm}=-(\frac{1}{2}\alpha \pm\sqrt{\Gamma _{L}\Gamma_{R}e^{2ik_{0}d}})$. The time-domain solutions are obtained by evaluating the inverse Laplace transforms via the residue theorem as
\begin{eqnarray}
\langle \hat{m}_{1}\rangle =-i\Omega \lbrack \frac{\frac{\alpha}{2} }{s_{+}s_{-}}+\frac{(s_{+}+\frac{\alpha}{2} )e^{s_{+}t}}{s_{+}(s_{+}-s_{-})}+\frac{(s_-+\frac{\alpha}{2} )e^{s_-t}}{s_-(s_--s_+)}],~~~\\
\langle \hat{m}_{2}\rangle=i\Omega \Gamma _{R}e^{ik_{0}d}[\frac{1}{s_+s_-}+\frac{s_+^{-1}e^{s_+t}}{s_+-s_-}+\frac{s_-^{-1}e^{s_-t}}{s_--s_+}].~~~
\end{eqnarray}
In the long-time condition, we have their steady-state form as
\begin{eqnarray}
\langle\hat{m}_1\rangle_\text{ss}&=&{-i\Omega\alpha/2\over s_+s_-}={-2i\Omega\alpha\over \alpha^2-4\Gamma_L\Gamma_Re^{2ik_0d}}, \\
\langle\hat{m}_2\rangle_\text{ss}&=&{i\Omega\Gamma_Re^{ik_0d}\over s_+s_-}={4i\Omega\Gamma_Re^{ik_0d}\over \alpha^2-4\Gamma_L\Gamma_Re^{2ik_0d}}. 
\end{eqnarray}

Next, we can prove that 
\begin{eqnarray}
\langle\hat{m}_{l}\hat{m}_{l'}\rangle&=&\langle\hat{m}_{l}\rangle\langle\hat{m}_{l'}\rangle, \qquad l,l'\in \{1,2\},\label{S4}\\
\langle\hat{m}_{l}^{\dagger}\hat{m}_{l}\rangle&=&|\langle \hat{m}_{l}\rangle|^2+\kappa\bar{n}\mathcal{R}_{l}(t), \label{S13}
\end{eqnarray}
\begin{proof}
Performing the Laplace transform to Eqs. \eqref{appm1p}, \eqref{appm2p}, and \eqref{appm12}, we obtain
\begin{eqnarray}
\mathcal{L}(\langle\hat{m}_{1}^2\rangle)&=& 
\frac{2i\Omega\Gamma_{L}e^{ik_{0}d}}{(s+\alpha)^{2}-4\Gamma_{L}\Gamma_{R}e^{2ik_{0}d}}[\mathcal{L}(\langle \hat{m}_{2}\rangle)\nonumber \\
&&-\frac{[(s+\alpha)^{2}-2\Gamma_{L}\Gamma_{R}e^{2ik_{0}d}]\mathcal{L}(\langle\hat{m}_{1}\rangle)}{\Gamma_{L}e^{ik_{0}d}(s+\alpha)}],~~~\\
\mathcal{L}(\langle\hat{m}_{2}^2\rangle)&=& {2i\Omega\Gamma_{R}e^{ik_{0}d}[\mathcal{L}(\langle \hat{m}_{2}\rangle)-\frac{2\Gamma_{R}\mathcal{L}(\langle\hat{m}_{1}\rangle)}{(s+\alpha)e^{-ik_{0}d}}]\over (s+\alpha)^{2}-4\Gamma_{L}\Gamma_{R}e^{2ik_{0}d} }.  ~~\label{S3}
\end{eqnarray}
On the other hand, we derive the equations of $\langle \hat{m}_l\rangle\frac{d\langle\hat{m}_{l'}\rangle}{dt}$, with $l,l'\in \{1,2\}$ from Eqs. \eqref{appm1} and \eqref{appm2}. Then from their Laplace transforms and the property $\mathcal{L}(f\frac{df}{dt})=\frac{s}{2}\mathcal{L}(f^{2})$, we obtain 
\begin{eqnarray}
\frac{s}{2}\mathcal{L}(\langle \hat{m}_{1}\rangle ^{2}) =-\frac{\alpha}{2}\mathcal{L}(\langle \hat{m}_{1}\rangle ^{2})-\Gamma _{L}e^{ik_{0}d}\mathcal{L}(\langle \hat{m}_{1}\rangle \langle \hat{m}_{2}\rangle )\nonumber\\
-i\Omega \mathcal{L}(\langle \hat{m}_{1}\rangle ),  ~~~~~~~~~~~~~~~~~~~~~~~~~~~~~~\label{S1} \\
\frac{s}{2}\mathcal{L}(\langle \hat{m}_{2}\rangle ^{2}) =-\frac{\alpha}{2}\mathcal{L}(\langle \hat{m}_{2}\rangle ^{2})-\Gamma _{R}e^{ik_{0}d}\mathcal{L}(\langle \hat{m}_{1}\rangle \langle \hat{m}_{2}\rangle ),  ~~~\\
s\mathcal{L}(\langle \hat{m}_{1}\rangle \langle \hat{m}_{2}\rangle )=-\alpha \mathcal{L}(\langle \hat{m}_{1}\rangle \langle \hat{m}_{2}\rangle)-\Gamma _{R}e^{ik_{0}d}\mathcal{L}(\langle \hat{m}_{1}\rangle ^{2})\nonumber\\-\Gamma
_{L}e^{ik_{0}d}\mathcal{L}(\langle \hat{m}_{2}\rangle ^{2})-i\Omega \mathcal{L}(\langle \hat{m}_{2}\rangle ), ~~~~ \label{S20}
\end{eqnarray}
They readily result in
\begin{eqnarray}
&&\mathcal{L}(\langle\hat{m}_{1}\rangle^2)=\frac{2i\Omega\Gamma_{L}e^{ik_{0}d}}{(s+\alpha)^{2}-4\Gamma_{L}\Gamma_{R}e^{2ik_{0}d}}[\mathcal{L}(\langle \hat{m}_{2}\rangle)\nonumber \\
&&~~~~~~~~~-\frac{[(s+\alpha)^{2}-2\Gamma_{L}\Gamma_{R}e^{2ik_{0}d}]\mathcal{L}(\langle\hat{m}_{1}\rangle)}{\Gamma_{L}e^{ik_{0}d}(s+\alpha)}],~~\\
&&\mathcal{L}(\langle \hat{m}_{2}\rangle ^{2})={2i\Omega\Gamma_{R}e^{ik_{0}d}[\mathcal{L}(\langle \hat{m}_{2}\rangle)-\frac{2\Gamma_{R}\mathcal{L}(\langle\hat{m}_{1}\rangle)}{(s+\alpha)e^{-ik_{0}d}}]\over (s+\alpha)^{2}-4\Gamma_{L}\Gamma_{R}e^{2ik_{0}d} },~~~~~~  \label{S2}
\end{eqnarray}
which matches Eq. \eqref{S3} exactly. Therefore, we have $\langle\hat{m}_{l}^2\rangle=\langle\hat{m}_{l}\rangle^{2}$ after the inverse Laplace transform. 

Performing the Laplace transform to Eq. \eqref{appm12}, we obtain
\begin{eqnarray}
s\mathcal{L}(\langle \hat{m}_{1}\hat{m}_{2}\rangle )&=&-\alpha \mathcal{L}(\langle \hat{m}_{1}\hat{m}_{2}\rangle)-\Gamma _{R}e^{ik_{0}d}\mathcal{L}(\langle \hat{m}_{1}^{2}\rangle )\nonumber\\
&&-\Gamma_{L}e^{ik_{0}d}\mathcal{L}(\langle \hat{m}_{2}^{2}\rangle)-i\Omega \mathcal{L}(\langle \hat{m}_{2}\rangle ), ~~~~ \label{S18}
\end{eqnarray}
The comparison of Eqs. (\ref{S20}) and (\ref{S18}) leads to
\begin{eqnarray}
\mathcal{L}(\langle\hat{m}_{1}\hat{m}_{2}\rangle)&=&\mathcal{L}(\langle\hat{m}_{1}\rangle\langle\hat{m}_{2}\rangle),  \label{S19}
\end{eqnarray}
where $\mathcal{L}(\langle\hat{m}_{l}^{2}\rangle)=\mathcal{L}(\langle\hat{m}_{l}\rangle^{2})$ has been used. Then, we have $\langle\hat{m}_{1}\hat{m}_{2}\rangle=\langle\hat{m}_{1}\rangle\langle\hat{m}_{2}\rangle$ after the inverse Laplace transform.

The Laplace transform converts Eqs. \eqref{apmm1}, \eqref{apmm2}, and \eqref{apm1m2} into
\begin{widetext}
\begin{eqnarray}
\mathcal{L}(\langle \hat{m}_{1}^{\dagger }\hat{m}_{1}\rangle ) &=&\frac{\Omega [i(s+\alpha )\mathcal{L}(\langle \hat{m}_{1}\rangle )-i\Gamma_{L}e^{ik_{0}d}\mathcal{L}(\langle \hat{m}_{2}\rangle )+\text{c.c.}]+2\Gamma _{L}^{2}\mathcal{L}(\langle \hat{m}_{2}^{\dagger }\hat{m}_{2}\rangle )+\frac{\kappa \bar{n}(s+\alpha )}{s}}{(s+\alpha )^{2}-2\Gamma _{L}\Gamma _{R}\cos (2k_{0}d)},  \label{S7}\\
\mathcal{L}(\langle \hat{m}_{2}^{\dagger }\hat{m}_{2}\rangle ) &=&\frac{\Omega \Gamma _{R}}{[(s+\alpha )^{2}-2\Gamma _{L}\Gamma _{R}\cos(2k_{0}d)]^{2}-4\Gamma _{L}^{2}\Gamma _{R}^{2}}\{[2\Gamma _{L}\Gamma _{R}\sin (2k_{0}d)-i(s+\alpha )^{2}]e^{-ik_{0}d}\mathcal{L}(\langle \hat{m}_{2}\rangle )\nonumber\\&&+2i\Gamma _{R}(s+\alpha )\mathcal{L}%
(\langle \hat{m}_{1}\rangle )+\text{c.c.}\} +\frac{\kappa \bar{n}(s+\alpha )[(s+\alpha )^{2}-2\Gamma _{L}\Gamma_{R}\cos (2k_{0}d)+2\Gamma _{R}^{2}]}{s\{[(s+\alpha )^{2}-2\Gamma _{L}\Gamma_{R}\cos (2k_{0}d)]^{2}-4\Gamma _{L}^{2}\Gamma _{R}^{2}\}}.  \label{S8}
\end{eqnarray}
On the other hand, from the Laplace transforms of Eqs. \eqref{appm1} and \eqref{appm2}, we have
\begin{eqnarray}
\mathcal{L}(| \langle \hat{m}_{1}\rangle|^2) &=&\frac{\Omega [i(s+\alpha )\mathcal{L}(\langle \hat{m}_{1}\rangle )-i\Gamma _{L}e^{ik_{0}d}\mathcal{L}(\langle \hat{m}_{2}\rangle )+\text{c.c.}]+2\Gamma _{L}^{2}\mathcal{L}(|\langle \hat{m}_{2}\rangle |^2)}{(s+\alpha )^{2}-2\Gamma _{L}\Gamma _{R}\cos (2k_{0}d)},  \label{S5}\\
\mathcal{L}(| \langle \hat{m}_{2}\rangle|^2) &=&\frac{\Omega \Gamma _{R}\{\left[ 2\Gamma _{L}\Gamma _{R}\sin (2k_{0}d)-i(s+\alpha )^{2}\right]
e^{-ik_{0}d}\mathcal{L}(\langle \hat{m}_{2}\rangle )+2i\Gamma _{R}(s+\alpha
)\mathcal{L}(\langle \hat{m}_{1}\rangle )+\text{c.c.}\}}{[(s+\alpha )^{2}-2\Gamma _{L}\Gamma _{R}\cos
(2k_{0}d)]^{2}-4\Gamma _{L}^{2}\Gamma _{R}^{2}}. \label{S6} 
\end{eqnarray}
\end{widetext}
The comparison of Eqs. (\ref{S7}), (\ref{S8}) and Eqs. (\ref{S5}), (\ref{S6}) leads to
\begin{eqnarray}
\mathcal{L}(\langle\hat{m}_{l}^{\dagger}\hat{m}_{l}\rangle)&=&\mathcal{L}(|\langle\hat{m}_{l}\rangle|^2)+\kappa\bar{n}\mathcal{T}_{l},  \label{S9}
\end{eqnarray}
with
\begin{eqnarray}
\mathcal{T}_{\text{1/2}}=\frac{(s+\alpha)[2\Gamma_{L/R}^{2}+(s+\alpha)^{2}-2\Gamma_{L}\Gamma_{R}\cos(2k_{0}d)]}{s\{
[(s+\alpha)^{2}-2\Gamma_{L}\Gamma_{R}\cos(2k_{0}d)]^{2}-4\Gamma_{L}^{2}\Gamma_{R}^{2}\}}.\nonumber \\ \label{S11}
\end{eqnarray}
Making the inverse Laplace transform to Eq. \eqref{S9}, we obtain Eq. \eqref{S13}, with $\mathcal{R}_l(t)$ being the inverse Laplace transform of $\mathcal{T}_l$. 
\end{proof}

In the following, we evaluate $\mathcal{R}_{l}(t)=\sum_{i=1}^{5}\text{Res}_{s=s_{i}}[\mathcal{T}_{l}(s)e^{st}]$.  The poles of $\mathcal{T}_{l}(s)$ are determined by $s\{[(s+\alpha)^{2}-2\Gamma_{L}\Gamma_{R}\cos(2k_{0}d)]^{2}-4\Gamma_{L}^{2}\Gamma_{R}^{2}\}=0$, which gives
\begin{eqnarray}
s_{1}&=&0,  \label{S17} \\
s_{2/3}&=&-\alpha+\sqrt{2\Gamma_{L}\Gamma_{R}[\cos(2k_{0}d)\pm1]},  \\
s_{4/5}&=&-\alpha-\sqrt{2\Gamma_{L}\Gamma_{R}[\cos(2k_{0}d)\pm1]}.   
\end{eqnarray}
According to the residue theorem, we obtain $\mathcal{R}_{l}(t)$ straightforwardly. Their steady-state solutions read
\begin{eqnarray}
\mathcal{R}_{1/2,\text{ss}}=\frac{\alpha[\alpha^{2}-2\Gamma_{L}\Gamma_{R}\cos(2k_{0}d)+2\Gamma_{L/R}^{2}]}{(\alpha^{2}-2\Gamma_{L}\Gamma_{R}\cos(2k_{0}d))^{2}-4\Gamma_{L}^{2}\Gamma_{R}^{2}}.
\end{eqnarray}

Next, we prove
\begin{eqnarray}
\langle\hat{m}_{1}^{\dagger}\hat{m}_{2}\rangle&=&\langle\hat{m}_{1}^{\dagger}\rangle\langle\hat{m}_{2}\rangle \nonumber\\
&&-\kappa\bar{n}[\Gamma_{R}e^{ik_{0}d}\mathcal{Q}_{1}(t)+\Gamma_{L}e^{-ik_{0}d}\mathcal{Q}_{2}(t)].~~~~~~~\label{S17}
\end{eqnarray}
\begin{proof}
The Laplace transform converts Eq. \eqref{apm1m2} into
\begin{eqnarray}
s\mathcal{L}(\langle \hat{m}_{1}^{\dagger}\hat{m}_{2}\rangle)=-\alpha\mathcal{L}(\langle \hat{m}_{1}^{\dagger}\hat{m}_{2}\rangle) -\Gamma_{R}e^{ik_{0}d}\mathcal{L}(\langle \hat{m}_{1}^{\dagger}\hat{m}_{1}\rangle)\nonumber\\
-\Gamma_{L}e^{-ik_{0}d}\mathcal{L}(\langle \hat{m}_{2}^{\dagger}\hat{m}_{2}\rangle) +i\Omega\mathcal{L}(\langle \hat{m}_{2}\rangle). ~~~~  \label{S21}
\end{eqnarray}
The substitution of Eq. \eqref{S9} into Eq. \eqref{S21} yields
\begin{eqnarray}
s\mathcal{L}(\langle \hat{m}_{1}^{\dagger}\hat{m}_{2}\rangle)&=&-\alpha\mathcal{L}(\langle \hat{m}_{1}^{\dagger}\hat{m}_{2}\rangle) -\Gamma_{R}e^{ik_{0}d}\mathcal{L}(|\langle\hat{m}_{1}\rangle|^2)\nonumber\\
&&-\Gamma_{L}e^{-ik_{0}d}\mathcal{L}(|\langle\hat{m}_{2}\rangle|^2) 
+i\Omega\mathcal{L}(\langle \hat{m}_{2}\rangle)\nonumber\\
&&-\kappa\bar{n}(\Gamma_{R}e^{ik_{0}d}\mathcal{T}_{1}+\Gamma_{L}e^{-ik_{0}d}\mathcal{T}_{2}). ~~~~  \label{S23}
\end{eqnarray}
On the other hand, by deriving the expressions for $\langle \hat{m}_{2}\rangle\frac{d\langle\hat{m}_{1}^{\dagger}\rangle}{dt}$ and $\langle \hat{m}_{1}^{\dagger}\rangle\frac{d\langle\hat{m}_{2}\rangle}{dt}$ from Eqs. \eqref{appm1} and \eqref{appm2}, we can expand the time derivative of the product $\langle\hat{m}_{1}^{\dagger}\rangle\langle \hat{m}_{2}\rangle$. Taking its Laplace transform, we obtain
\begin{eqnarray}
s\mathcal{L}(\langle\hat{m}_{1}^{\dagger}\rangle\langle \hat{m}_{2}\rangle)=-\alpha\mathcal{L}(\langle\hat{m}_{1}^{\dagger}\rangle\langle \hat{m}_{2}\rangle)-\Gamma_{R}e^{ik_{0}d}\mathcal{L}(|\langle \hat{m}_{1}\rangle|^2)\nonumber\\
-\Gamma_{L}e^{-ik_{0}d}\mathcal{L}(|\langle \hat{m}_{2}\rangle|^2) +i\Omega\mathcal{L}(\langle \hat{m}_{2}\rangle). ~~~~  \label{S22}
\end{eqnarray}
Comparing Eq. \eqref{S23} with Eq. \eqref{S22}, we find
\begin{eqnarray}
\mathcal{L}(\langle \hat{m}_{1}^{\dagger}\hat{m}_{2}\rangle)&=&\mathcal{L}(\langle \hat{m}_{1}^{\dagger}\rangle\langle\hat{m}_{2}\rangle)\nonumber\\
&&-{\kappa\bar{n}\over s+\alpha}(\Gamma_{R}e^{ik_{0}d}\mathcal{T}_{1}+\Gamma_{L}e^{-ik_{0}d}\mathcal{T}_{2}).~~~~~~
\end{eqnarray}
Finally, we arrive at Eq. \eqref{S17} after the inverse Laplace transform, where $\mathcal{Q}_{l}(t)=\sum_{i=1}^{5}\text{Res}_{s=s_{i}}[\frac{\mathcal{T}_{l}(s)}{s+\alpha}e^{st}]$.
\end{proof}

\section{Energies, ergotropy, and charger-QB entanglement}\label{Appendix:Energies and ergotropy}

The elements of the covariance matrix ${\pmb\sigma}^{(j)}(t)$ and the mean vector ${\bf d}^{(j)}(t)$ of the charger for $j=1$ and the QB for $j=2$ are $\sigma^{(j)}_{lm}=\text{Tr}[\{\Delta \hat{r}_{l},\Delta \hat{r}_{m}\}\rho_j(t)]$ and $d^{(j)}_{l}=\text{Tr}[\hat{r}^{(j)}_{l}\rho_j(t)]$, with $\Delta\hat{r}^{(j)}_{l}=\hat{r}^{(j)}_{l}-d^{(j)}_{l}$ and $\hat{r}^{(j)}=({\hat{m}_{j}^{\dagger}+\hat{m}_{j}\over\sqrt{2}},{\hat{m}_{j}-\hat{m}_{j}^{\dagger}\over i\sqrt{2}})^{\text{T}}$. Thus, we have 
\begin{eqnarray}
d^{(j)}_1(t)&=&{\langle \hat{m}_{j}\rangle/\sqrt{2}}+\text{c.c.},\\
d^{(j)}_2(t)&=&{\langle \hat{m}_{j}\rangle/( i\sqrt{2})}+\text{c.c.},\\
\sigma^{(j)} _{11}(t) &=&\langle \hat{m}_{j}^2\rangle +\langle \hat{m}_{j}^{\dagger2}\rangle +2\langle \hat{m}_{j}^{\dagger}\hat{m}_{j}\rangle +1\nonumber\\
&&-\langle \hat{m}_{j}\rangle ^2-2|\langle \hat{m}_{j}\rangle|^2 -\langle \hat{m}_{j}^{\dagger }\rangle^2\nonumber\\
&=&2\kappa \bar{n}\mathcal{R}_j(t)+1,\label{S31} \\
\sigma^{(j)} _{12}(t) &=&\frac{1}{i}[\langle \hat{m}_{j}^2\rangle -\langle\hat{m}_{j}^{\dagger2}\rangle -\langle \hat{m}_{j}\rangle ^2 +\langle \hat{m}_{j}^{\dagger}\rangle^2 ]=0,   \\
\sigma^{(j)} _{22}(t) &=&-\langle \hat{m}_{j}^2\rangle -\langle \hat{m}_{j}^{\dagger2}\rangle +2\langle \hat{m}_{j}^{\dagger}\hat{m}_{j}\rangle +1\nonumber\\
&&+\langle \hat{m}_{j}\rangle^2-2| \langle \hat{m}_{j}\rangle |^2+\langle\hat{m}_{j}^{\dagger }\rangle^2 \nonumber\\
&=&2\kappa \bar{n}\mathcal{R}_j(t)+1,
\end{eqnarray}
where Eqs. \eqref{S4} and \eqref{S13} have been used. Substituting them into the expressions for the energy of the charger, the energy of the QB, and the ergotropy of the QB, we obtain
\begin{eqnarray}
\mathcal{E}_\text{C/B}(t)&=&\omega_0\{{\text{Tr}[{\pmb\sigma^{(1/2)}}(t)]/ 4}+{|{\bf d}^{(1/2)}(t)|^2/ 2}\}\nonumber\\
&=&\omega_0[|\langle \hat{m}_{1/2}\rangle|^2+\kappa\bar{n}\mathcal{R}_{1/2}(t)+{1\over 2}],\label{eddcf}\\ 
\mathcal{W}(t)&=&\mathcal{E}_\text{B}(t)-\omega_0{\sqrt{\text{det}[{\pmb\sigma}^{(2)}(t)]}}/{2}=\omega_0|\langle \hat{m}_2\rangle|^2.~~~~~~
\end{eqnarray}
At zero temperature, Eq. \eqref{eddcf} reduces to $\mathcal{E}_{\text{C/B}}(\infty)=\omega_{0}(|\langle\hat{m}_{1/2}\rangle_\text{ss}|^2+\frac{1}{2})$ in the long-time condition.

The entanglement of a continuous-variable Gaussian state is measured by the logarithmic negativity \cite{PhysRevA.65.032314,PhysRevLett.95.090503} 
\begin{equation}
E_N = \max\bigl\{0, -\log_2[\min(\tilde{\nu}_{+},\tilde{\nu}_{-})] \bigr\},
\end{equation}
where $\tilde{\nu}_{\pm}$ are the symplectic eigenvalues of the partially transposed covariance matrix. $\tilde{\nu}_{\pm}$ relate to the covariance matrix ${\pmb\sigma}(t) = \begin{pmatrix}
\pmb\sigma^{(1)}(t) & \mathcal{C} (t)\\
\mathcal{C}^{\text{T}}(t) & \pmb\sigma^{(2)}(t)
\end{pmatrix}$ as
\begin{equation}
\tilde{\nu}_{\pm}= \sqrt{\big[\delta \pm \sqrt{\delta^2 - 4\det {\pmb\sigma}}\,\big]/2},
\end{equation}
where the partial-transposition invariant $\delta$ can be simplified to $\delta = \det \pmb\sigma^{(1)}(t) + \det \pmb\sigma^{(2)}(t) - 2\det \mathcal{C}(t)$. Using Eqs. \eqref{S4}, \eqref{S13}, and  \eqref{S17} at zero temperature $\bar{n}=0$, we can evaluate 
\begin{eqnarray}
\det \mathcal{C}(t)=0,~\det \pmb\sigma^{(j)}(t) =\det \pmb\sigma(t) = 1.
\end{eqnarray}
They readily lead to $\tilde{\nu}_{\pm}= 1$ and thus $E_N = 0$. Therefore, the quantum resource responsible for the enhancement of the energy and ergotropy is not the charger-QB entanglement. The enhanced charging performance is instead understood in terms of the directional energy transfer and phase-sensitive interference induced by the chiral coupling.

\begin{figure}[t!]
\includegraphics[width=0.95\columnwidth]{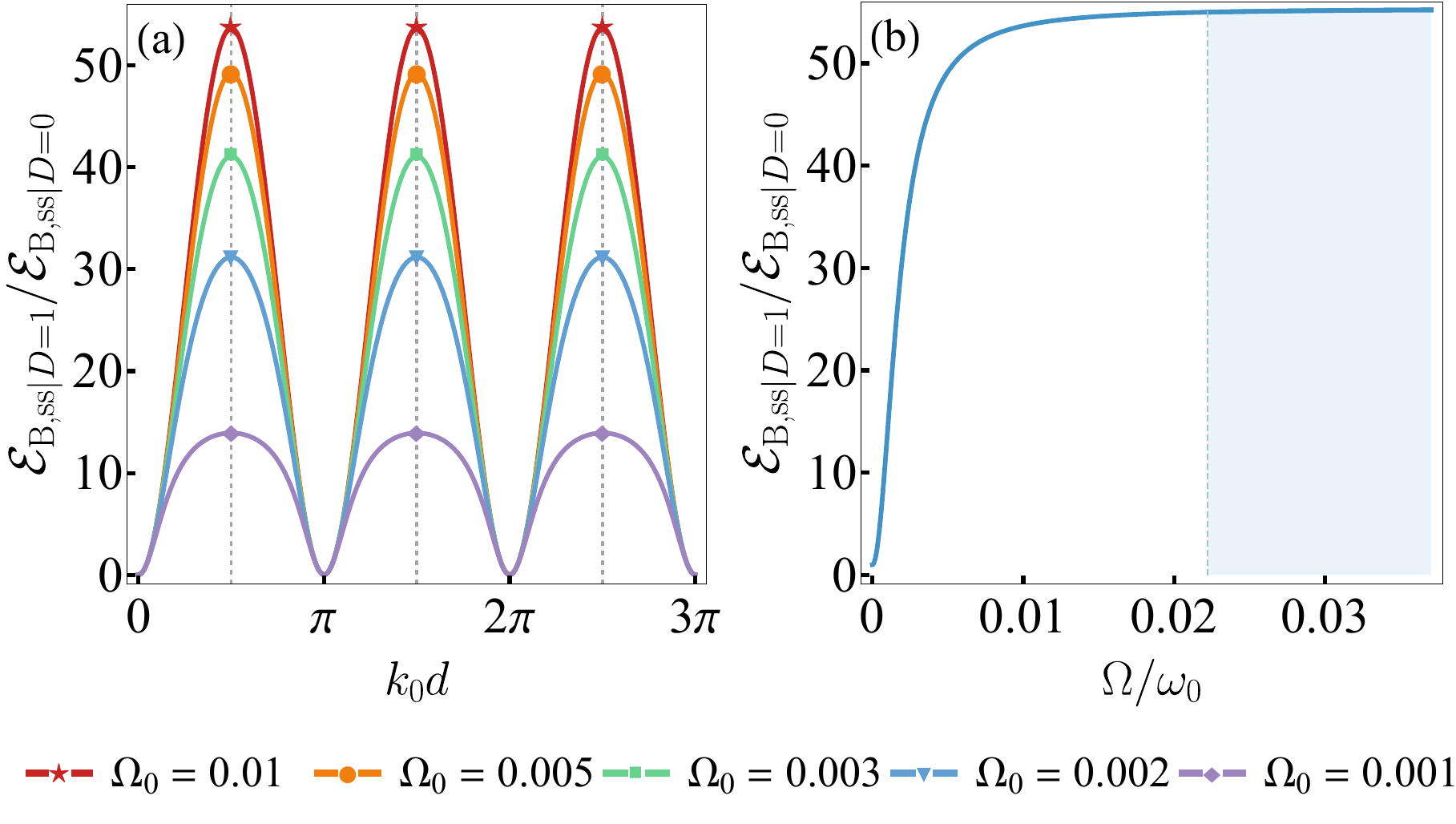}
\caption{(a) $\mathcal{E}_{\text{B,ss}|D=1}/\mathcal{E}_{\text{B,ss}|D=0}$ as a function of the normalized distance $k_{0}d$ for different driving strengths. (b) $\mathcal{E}_{\text{B,ss}|D=1}/\mathcal{E}_{\text{B,ss}|D=0}$ versus driving strength at a fixed distance $k_{0}d=\pi/2$. The parameters are $\Gamma_R\approx0.001\omega_0$, $\bar{n}=0$, $\kappa=\frac{\Gamma_R}{20}$, and $\Omega_{0}=\Omega/\omega_{0}$.}\label{energy10}
\end{figure}

\begin{figure}[tbp]
\includegraphics[width=0.9\columnwidth]{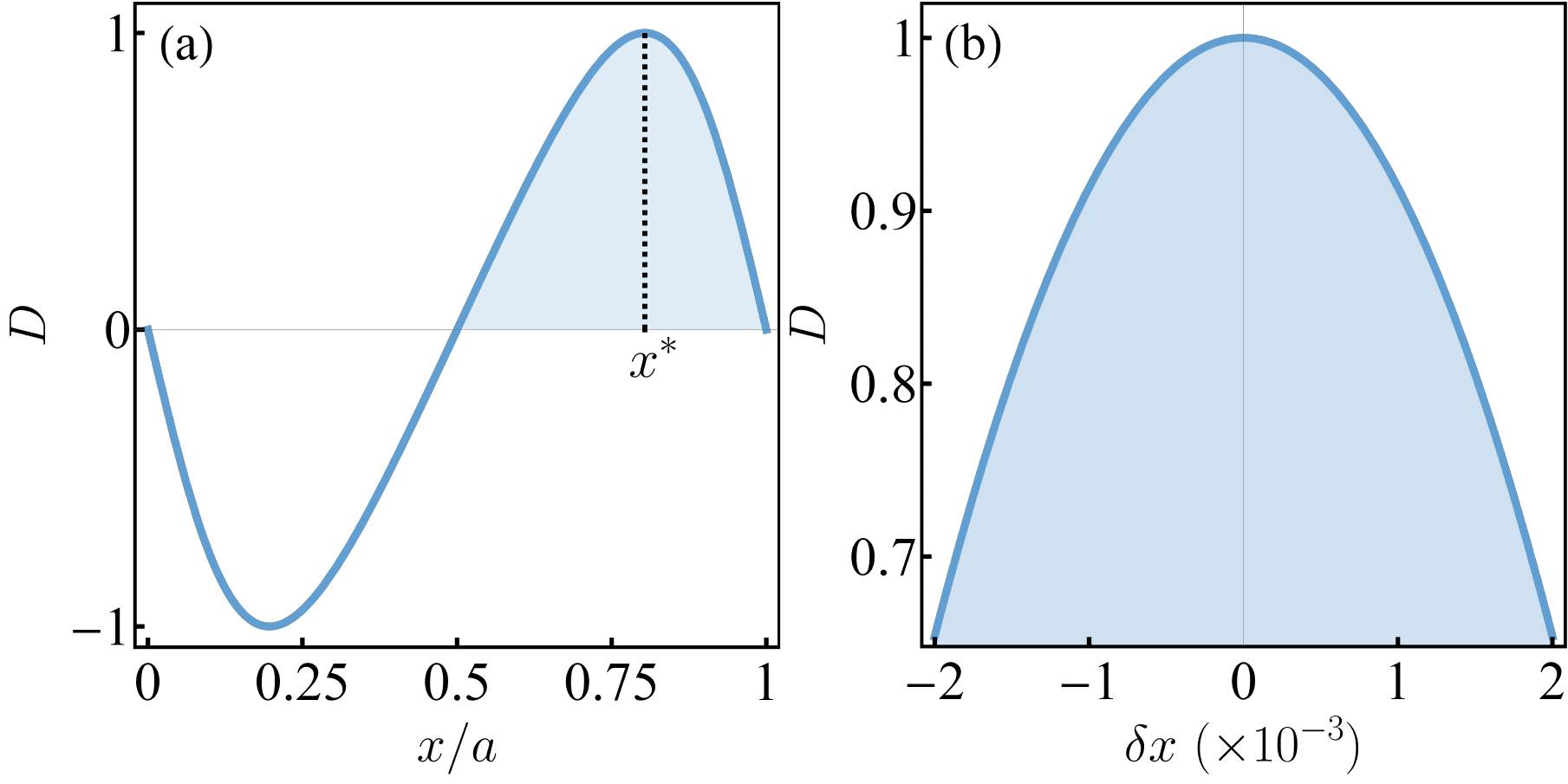}
\caption{(a) Chirality $D$ versus position $x$.  (b) $D$ versus displacement $\delta x$ from the optimal point $x^{\ast}$. Other parameters are the same as in Fig. 2 of the main text.}\label{chiralityx}
\end{figure}

\section{Effect of the driving strength}\label{Appendix:driving strength}
In the main text, our results show an enhancement factor of approximately 34 for $\mathcal{E}_{\text{B,ss}|D=1}/\mathcal{E}_{\text{B,ss}|D=0}$. Here we further demonstrate that increasing the driving strength raises this enhancement factor until it saturates at around 55. Figure~\ref{energy10}(a) plots the $\mathcal{E}_{\text{B,ss}|D=1}/\mathcal{E}_{\text{B,ss}|D=0}$ as a function of distance $k_{0}d$ under different driving strengths $\Omega$. The $\mathcal{E}_{\text{B,ss}|D=1}/\mathcal{E}_{\text{B,ss}|D=0}$ increases with the driving strength and exhibits a periodic dependence on $k_{0}d$, reaching its maximum at $d=(n+1/2)\pi/k_{0}$. Figure~\ref{energy10}(b) shows the variation of the $\mathcal{E}_{\text{B,ss}|D=1}/\mathcal{E}_{\text{B,ss}|D=0}$ with the driving strength at the fixed distance $d=\pi/2k_{0}$. The $\mathcal{E}_{\text{B,ss}|D=1}/\mathcal{E}_{\text{B,ss}|D=0}$ grows with the driving strength and eventually saturates at an enhancement factor of approximately 55. This indicates that the energy enhancement can be improved by increasing the driving strength.

\section{Sensitivity to experimental imperfections}\label{impftc}
Considering two identical YIG spheres with the same position ($x_1=x_2=x$) and equal volumes ($V_1=V_2=V$), the coupling strength mediated by TE$_{10}$ photons in our rectangular waveguide is
\begin{equation}
g_{\lambda}(\omega) =\sqrt{\frac{\gamma MV}{2\epsilon
_{0}\omega ab}}\left[ \frac{\pi }{a}\cos (\frac{\pi x}{a})
-k_{\lambda }\sin (\frac{\pi x}{a}) \right], 
\end{equation}
where $a=1.6$ cm and $b=0.7$ cm are the cross-sectional dimensions of the waveguide, and $k=\sqrt{\omega_{0}^{2}/c^{2}-\pi^{2}/a^{2}}$ is the positive wave vector. The relation $k_{R}=-k_{L}\equiv k$ gives $g_{L}(\omega)\neq g_{R}(\omega)$, which is the origin of chiral magnon-waveguide coupling.

We now examine how the position $x$ affects the nonreciprocity parameter $D$. The degree of nonreciprocity is
\begin{eqnarray}
D(x) &=&\frac{\Gamma _{R}(x) -\Gamma _{L}(x) }{\Gamma _{R}(x) +\Gamma _{L}(x)}, \nonumber \\ 
&=&-\frac{2\pi}{a}\frac{k\cos ( \frac{\pi x}{a}) \sin ( 
\frac{\pi x}{a})}{\left[(\frac{\pi }{a}) ^{2}\cos
^{2}(\frac{\pi x}{a})+k^{2}\sin^{2}(\frac{\pi x}{a})\right]}.\label{chiralityDx}
\end{eqnarray}
Figure~\ref{chiralityx}(a) shows the dependence of $D(x)$ on $x$. In this work, we focus on $D\in(0,1)$. The maximum $D(x)=1$ occurs at $x=x^{\ast}$. By solving $D(x^{\ast})=1$, we obtain
\begin{equation}
x^{\ast}=-\frac{a}{\pi}\arctan (\frac{\pi}{ak}) +an,\qquad n\in\mathbb{Z}.
\end{equation}
Choosing $n=1$ gives $x^{\ast}=a-\frac{a}{\pi}\arctan (\frac{\pi}{ak})$, which we take as the optimal position. Figure~\ref{chiralityx}(b) plots $D$ versus the offset $\delta x=x^{\ast}-x$ near this optimal point. Expanding around $x^{\ast}$ to the second order gives
\begin{equation}
D(x^{\ast}-\delta x) =1-\frac{(\pi ^{2}+(ak)
^{2})^{2}}{2a^{4}k^{2}}(\delta x)^{2}+O(\delta
x^{3}).\label{Dtaylorx}
\end{equation}
Even when $\delta x$ is as large as $1$ mm, which is 4 times the diameter of a typical $250\text{-}\mu\text{m}$ yttrium iron garnet sphere \cite{PhysRevLett.128.047701}, the parameter $D$ is still maintained at $0.91$. It indeed causes the degradation of the chirality enhancement effects. However, the purely quadratic scaling of the deviation ensures the absence of any sharp threshold for abrupt disappearance of the chiral advantage. Therefore, the enhancement effects caused by the position imperfection degrade smoothly instead of abruptly because the steady-state energy and ergotropy are analytic and continuous functions of $D$.

\bibliography{References}
\end{document}